\documentclass{article}

\usepackage{arxiv}

\usepackage[utf8]{inputenc} 
\usepackage[T1]{fontenc}    
\usepackage{lmodern}        
\usepackage{hyperref}       
\usepackage{url}            
\usepackage{booktabs}       
\usepackage{amsfonts}       
\usepackage{nicefrac}       
\usepackage{microtype}      
\usepackage{lipsum}
\usepackage{graphicx}

\title{Embeddings and Attention in Predictive Modeling}

\author{
    Kevin Kuo
   \\
    RStudio and Kasa AI \\
   \\
  \texttt{\href{mailto:kevinykuo@gmail.com}{\nolinkurl{kevinykuo@gmail.com}}} \\
   \And
    Ronald Richman
   \\
    QED Actuaries and Consultants and University of the Witwatersrand \\
   \\
  \texttt{\href{mailto:ron@ronaldrichman.co.za}{\nolinkurl{ron@ronaldrichman.co.za}}} \\
  }

\newlength{\csllabelwidth}
\newlength{\cslhangindent}
  {}%
  {\par}
\newenvironment{CSLReferences}[3] 
 {
  \setlength{\parindent}{0pt}
  \ifodd #1 \everypar{\setlength{\hangindent}{\cslhangindent}}\ignorespaces\fi
  \ifnum #2 > 0
  \setlength{\parskip}{#2\baselineskip}
  \fi
 }%
 {}
\usepackage{calc} 

\usepackage{amsmath}
\usepackage{mathtools}
\usepackage{bm}
\usepackage{bbm}
\usepackage{float}
\DeclarePairedDelimiter\ceil{\lceil}{\rceil}

\begin{document}
\maketitle

\def\tightlist{}

\begin{abstract}
We explore in depth how categorical data can be
processed with embeddings in the context of claim severity modeling. We develop several models that range
in complexity from simple neural networks to state-of-the-art attention based architectures that utilize embeddings.
We illustrate the utility of learned embeddings from neural networks as pretrained features in
generalized linear models, and discuss methods for visualizing and interpreting embeddings. Finally, we explore
how attention based models can contextually augment embeddings, leading to enhanced predictive performance.
\end{abstract}


\hypertarget{introduction}{%
\section{Introduction}\label{introduction}}

Categorical data are modelled by actuaries in many different contexts, from pricing of insurance products to reserving for the liabilities generated by these products. Sometimes, these data are modelled in an explicit manner, for example, when building models that apply across multiple categories, a form of dummy coding is usually used. For example, when building models to price the frequency of motor insurance claims, claim experience relating to different types of motor vehicle will often be modelled by including a (single) factor within models that modifies the relative frequency predicted for each type of vehicle. In other cases, modeling is performed for each category separately, thus the categorical data is used within the modeling in an implicit manner, for example, common practice is to estimate reserves for different lines of business separately, meaning to say, with no parameters being shared across each of the reserving models.

Recently, several studies of insurance problems have applied an alternative approach, originally from the natural known as \textbf{categorical embeddings}. Instead of trying to capture the differences between categories using a single factor, the categorical embedding approach rather maps each category to a low-dimensional numeric vector, which is then used within the model as a new predictor variable.

This approach to modeling categorical data has several advantages over more traditional treatments of categorical data. Using categorical embeddings instead of traditional techniques has been shown to increase predictive accuracy of models, for example, see Richman (2018) in the context of pricing. Models incorporating categorical embeddings can be pre-calibrated to traditional actuarial models, increasing the speed with which these models can be calibrated and leading to models with better explainability (Wüthrich and Merz 2019). Finally, the similarity between the vectors learned for different categories can be inspected, sometimes leading to insights into the workings of models, see, for example, Kuo (2019).

On the other hand, several open questions about the use of embeddings within actuarial work remain, which we aim to address in this study. First, hyperparameter settings for embeddings, such as the dimensions of the embedding layer and the use of regularization techniques such as dropout or normalization, that achieve optimal predictive performance has not yet been studied in detail in the actuarial literature. In this work, we aim to study how embeddings using different settings perform in the context of a large-scale predictive modeling problem, and give guidance on the process that can be followed to determine this in other problems. Although neural network have been shown to achieve excellent predictive accuracy on actuarial tasks, many actuaries still prefer to use GLM models for pricing tasks, thus, the issue of whether transferring embeddings to GLM models can achieve better performance is considered in this paper. Finally, in the past several years, a new type of neural network architecture based on attention (Vaswani et al. 2017a) has been successfully used on embeddings in the field of natural language processing and we incorporate attention based models into our predictive modeling example to demonstrate the use of these models for actuarial purposes.

In this work, we utilize the recently released National Flood Insurance Program (NFIP) dataset (Federal Emergency Management Agency 2019) which provides exposure information for policies written under the NFIP, as well as the claims data relating to these exposures.

The rest of this manuscript is organized as follows. Section \ref{lit_review} reviews recent applications of embeddings in the actuarial literature. Section \ref{defns} provides the notation used in the paper and defines GLMs, neural networks and related modeling concepts, including embeddings and attention. In Section \ref{modeling}, we provide initial models for the NFIP dataset and consider how successfully the embeddings used in the neural network model can be transferred to a GLM model of the same data and how learned embeddings can be interpreted. We focus on attention based models in Section \ref{attention_models}. Finally, Section \ref{Conclusions} provides a discussion of the results of this paper and considers avenues for future research.

\hypertarget{lit_review}{%
\section{Literature Review}\label{lit_review}}

Categorical data are usually modelled within GLMs and other predictive models using indicator variables which capture the effect of each level of the category, see, for example, Section 2 in Goldburd et al. (2020), using one of two main encoding schemes: dummy-coding and one-hot encoding. Dummy-coding, used in the popular \texttt{R} statistical software, assigns one level of the category as a baseline, for which an indicator variable is not calibrated, and the rest of the levels are assigned indicator variables, thus, producing estimates within the model of how the effects of each level differ from the baselines. One-hot encoding, often used in machine learning, is similar to dummy-coding, but assigns indicator variables to each level, in other words, calibrates an extra indicator variable compared with dummy-coding.

A different approach to modeling categorical data is credibility theory (see Bühlmann and Gisler (2005) for an overview), which, in the context of rating, can be applied to derive premiums that reflect the experience of a particular policyholder, by estimating premiums as a weighted average between the premium produced using the collective experience (i.e.~of all policyholders) and the premium produced using the experience of the particular policyholder. The weight used in this average is called a credibility factor and is calculated with reference to te variability of the policyholder experience relative to the variability of the group experience. In this context, the implicit categorical variable is the policyholder under consideration.

Generalized Linear Mixed Models (GLMMs) are an extension of GLMs that are designed for modeling categorical data using a principle very similar to that of credibility theory (Klinker 2010). Instead of calibrating indicator variables for each level of the category, GLMMs estimate effects for each of these levels as a combination of the overall group mean and the experience in each level of the category.

Embedding layers represent a different approach to the problem of modeling categorical data that was recently introduced in an actuarial context. Note that in the next section, we reflect on similarities between the conventional approaches discussed above and embedding layers. Richman (2018) reviewed the concept of embedding layers and connected the sharing of information across categories to the familiar concept of credibility theory. In that work, two applications of embedding layers were demonstrated. The first of these was in a Property and Casualty (P\&C) pricing context, it was shown that the out-of-sample accuracy of a neural network trained to predict claims frequencies on motor third party liability was enhanced by modeling the categorical variables within this dataset using embedding layers. Second, a neural network with embedding layers was used to model all of the mortality rates in the Human Mortality Database, where the differences in population mortality across countries and the differences in mortality at different ages were modelled with embedding layers, again producing more accurate out of sample performance than the other models tested.

Contemporaneous with that work is the DeepTriangle model of Kuo (2019), which applied recurrent neural networks to the problem of Incurred but not Reported (IBNR) loss reserving, to model jointly the paid and incurred losses in the NAIC Schedule P dataset. Embedding layers were used to capture the effect of differences in reserving delays and loss ratios for each company in the Schedule P dataset.

Many other applications of embeddings have subsequently appeared in the actuarial literature. Within mortality forecasting, Richman and Wüthrich (2019) and Perla et al. (2020) both apply embeddings layers to model and forecast mortality rates on a large scale. Wüthrich and Merz (2019) discussed how embeddings can be calibrated using GLM techniques and then incorporated into a combined actuarial neural network, with subsequent contributions in P\&C pricing by Schelldorfer and Wüthrich (2019) and in IBNR reserving by Gabrielli (2019) and Gabrielli, Richman, and Wüthrich (2019). Other applications in IBNR reserving are in Kuo (2020) and Delong, Lindholm, and Wuthrich (2020) who use embedding layers to model individual claims development.

\hypertarget{defns}{%
\section{Theoretical Overview}\label{defns}}

In this study, we are concerned with regression modeling, which is the task of predicting an unknown outcome \(y\) on the basis of information about that outcome contained in predictor variables, or features, stored in a matrix \(X\). For simplicity, we only consider the case of univariate outcomes, i.e., \(y \in \mathbb{R}^1\). The outcomes and the rows of the predictor variable matrix are indexed by \(i \in \{1 \dots I\}\), where \(i\) represents a particular observation of \((y_i, \mathbf{x}_i)\), where bold indicates that we are now dealing with a vector. The columns of the predictor variables are indexed by \(j \in \{1 \dots J\}\), where \(j\) represents a particular predictor variable, of which \(J\) have been observed, thus, we use the notation \(X_j\) to represent the \(j\)th predictor variable and \(X \in \mathbb{R}^J\). Formally, we look to build regression models that map from the predictor variables \(\mathbf{x}_i\) to the outcome \(y\) using a function \(f\) of the form:

\[
\mathbf{f} : \mathbb{R}^J \mapsto \mathbb{R}^1, \quad \quad \mathbf{x_i} \mapsto  \mathbf f{(\mathbf{x_i})} = y. 
\]

In this study, will use mainly use GLMs and neural networks to approximate the function \(f(.)\).

The predictor variables that we consider here are comprised of two types: continuous variables, taking on numerical values and represented by the matrix \(X_{num}\) with \(|J^{num}|\) columns, and categorical variables, which take on discrete values indicating one of several possible categories, represented by the matrix \(X_{cat}\) with \(|J^{cat}|\) columns, such that \(|J^{num}| + |J^{cat}| = J\).

\hypertarget{catdata}{%
\subsection{Categorical data modeling}\label{catdata}}

A categorical variable \(X_j, j \in J^{cat}\) takes as its value only one of a finite number of labels. Let the set of labels be \(\mathcal{P}^j = \{p^j_1, p^j_2,\dots, p^j_{n_{\mathcal{P}^j}} \}\), where \(n_\mathcal{P}^j = |\mathcal{P}^j|\) is the cardinality or number of levels, in \(\mathcal{P}^j\). One-hot encoding maps each value \(x_{i,j}\) of \(X_j\) to \(n_\mathcal{P}^j\) indicator variables, which take a value of \(1\) if the label of \(x_{i,j}\) corresponds to the level of the indicator variable, and \(0\) otherwise. An example of one-hot encoding is shown in Table \ref{tab:onehot}.

\begin{table}

\caption{\label{tab:onehot}Example one-hot encoding of the state variable}
\centering
\begin{tabular}[t]{l|r|r|r|r|r}
\hline
state & state\_CA & state\_MD & state\_ND & state\_UT & state\_WA\\
\hline
CA & 1 & 0 & 0 & 0 & 0\\
\hline
MD & 0 & 1 & 0 & 0 & 0\\
\hline
ND & 0 & 0 & 1 & 0 & 0\\
\hline
UT & 0 & 0 & 0 & 1 & 0\\
\hline
WA & 0 & 0 & 0 & 0 & 1\\
\hline
\end{tabular}
\end{table}

One-hot encoding is often used in the machine learning community while the statistical community often favors dummy coding, which, instead of assigning \(n_p^j\) indicator variables, assigns one of the levels of the categories as a baseline, and maps all of the other \(n_p^j -1\) variables to indicator variables. An example of dummy encoding is shown in Table \ref{tab:dummy}.

\begin{table}

\caption{\label{tab:dummy}Example dummy encoding of the state variable}
\centering
\begin{tabular}[t]{l|r|r|r|r}
\hline
state & state\_MD & state\_ND & state\_UT & state\_WA\\
\hline
CA & 0 & 0 & 0 & 0\\
\hline
MD & 1 & 0 & 0 & 0\\
\hline
ND & 0 & 1 & 0 & 0\\
\hline
UT & 0 & 0 & 1 & 0\\
\hline
WA & 0 & 0 & 0 & 1\\
\hline
\end{tabular}
\end{table}

After encoding the categorical data in this manner, most regression models such as GLMs will then fit coefficients for each level of the category in the table (if a tree based model is used, such as decision tree, then splits in the tree may occur depending on the presence, or not, of the categorical variable for the data). If one-hot encoding has been used, \(n_\mathcal{P}^j\) coefficients will be fit, compared to \(n_\mathcal{P}^j - 1\) coefficients in the case of dummy coding.

These coefficients represent the effect that each level of the categorical variable will have on the outcome. In the case that there are no other variables available in the dataset, then the coefficients will reflect the average value of the outcomes for that level of the categorical variable. For example, suppose that the categorical variable is a policyholder identifier, and the outcomes are the value of claims in different years, then the coefficients will reflect the average annual claims for each policyholder based on the experience. In other words, both of these encoding schemes give full credibility to the data available for each category, thus, even if a relatively small amount of data is available for a specific policyholder, the coefficient that is calibrated will only reflect that data. On the other hand, a foundational technique within actuarial work is the application of credibility methods, which are used for experience rating and other applications. These techniques provide an estimate that reflects not only the experience of the individual policyholder but also that of the collective, based on an estimate of how credible the data for each individual is. While we have described the application of credibility in a simple univariate context, it is also possible to apply credibility considerations within GLMs. using models known as Generalized Linear Mixed Models or GLMMs, and we refer to Klinker (2010) for more details.

Having described traditional approaches for modeling categorical data, we now turn to neural networks, and discuss embedding layers for categorical data modeling, which we define in more detail in the section on neural networks.

\hypertarget{neural-networks}{%
\subsection{Neural Networks}\label{neural-networks}}

Neural networks are flexible machine learning models that have recently been applied to a number of problems with Property and Casualty (P\&C) insurance. Here, we provide a brief overview of these models, and refer the reader to Richman (2018) for a more detailed overview. Neural networks are characterized by multiple layers of non-linear regression functions that are used to learn a new representation of the data input to the network that is then used to make predictions. Here we focus on the most common type of neural networks, which are fully connected networks (FCNs), which provide as the output of each set of non-linear functions to the subsequent layer of functions. Formally, a \(K\)-layer neural network is:

\begin{equation} 
\label{NN}
\begin{split}
z^1 &= \sigma(a_1.X+ b_1) \\
z^2 &= \sigma(a_2.z^1+ b_2) \\
\vdots \\
z^K &= \sigma(a_{K}.z^{K-1}+ b_{K}) \\
\hat{y} &= \sigma(a_{K+1}.z^{K}+ b_{K+1}),
\end{split}
\end{equation}

where the regression parameters (weights) for each layer \(k \in [1;K+1]\) are represented by the matrices \(a_k\) and the intercept terms are represented by \(b_k\). Whereas the calculation inside each of the layers is nothing more than linear regression, \(\sigma\) represents the non-linear part of each layer. Choices for \(\sigma\) are often the hyperbolic tangent (tanh) function or the rectified linear unit (ReLu) \(max(0,x)\). The parameters of the network are estimated (`trained') as follows. First a loss function \(L(.,.)\) is specified for the network that measures the difference between the observed data \(y\) and the predictions of the network \(\hat{y}\), for example, the Mean Squared Error (\((y - \hat{y})^2\)). Then, the parameters of the network are changed such that the loss decreases (formally, this is done using the technique of backpropagation). Finally, training is stopped once the predictive performance of the network on unseen data is suitably good.

If \(K\) is set equal to \(1\), then Equation \ref{NN} reduces to nothing more than a GLM. A neural network with \(K=2\) is called a shallow neural network and for \(K \geq 2\), the network is called a deep neural network. The matrix \(X\) of data input to the network can be composed of both continuous variables as well as categorical variables, which can be pre-processed using one-hot or dummy encoding. As mentioned above, a different option is to use encodings, which we discuss in more detail next.

\hypertarget{embeddings}{%
\subsubsection{Embeddings}\label{embeddings}}

Common issues with the traditional encoding schemes for categorical data occur when the number of levels for each variable is very large. Often, in these cases, models do not converge quickly, and the very large matrices that result from applying these schemes often cause computational difficulties. Besides for these practical issues, a deeper issue is that one-hot or dummy encoded data assumes that each category is entirely independent of the rest of the categories, in other words, there are no similarities between categories that could enable more robust estimation of models. In technical terms, this is because the columns of the matrices created by one-hot encoding are all orthogonal to each other. (These arguments appear in a similar form in Guo and Berkhahn (2016).) Solutions to these problems are provided by embedding layers.

An embedding layer is a neural network component which maps each level of the categorical data to a low dimensional vector of parameters that is learned together with the rest of the GLM or neural network that is used for the modeling problem. Formally, an embedding is

\begin{equation} 
\label{embed_map}
  z_\mathcal{P}^j : \mathcal{P}^j \to \mathbb{R}^{q_\mathcal{P}^j }, \quad \quad p^j \mapsto  z_\mathcal{P}^j(p),
\end{equation}

where \(q_\mathcal{P}^j\) is the dimension of the embedding for the \(j\)th categorical variable and \(z_\mathcal{P}^j(.)\) is an implicit function that maps from the particular element of the labels \(p\) to the embedding space. Equation \ref{embed_map} states that an embedding maps a level of a categorical variable to a numerical vector. This function is left implicit, meaning to say, we allow the embeddings to be derived during the process of fitting the model and do not attempt to specify exactly how the embeddings can be derived from the input data. In Table \ref{tab:embed} we show an example of two dimensional embeddings for the state variables, where these have been generated randomly.

\begin{table}

\caption{\label{tab:embed}Example (random) embeddings of the state variable}
\centering
\begin{tabular}[t]{l|r|r}
\hline
state & dimension1 & dimension2\\
\hline
CO & 1.5115220 & 2.2866454\\
\hline
DC & -0.0946590 & -1.3888607\\
\hline
ME & 2.0184237 & -0.2787888\\
\hline
PA & -0.0627141 & -0.1333213\\
\hline
UT & 1.3048697 & 0.6359504\\
\hline
\end{tabular}
\end{table}

When applying embeddings in a data modeling context using neural networks, the values of the embeddings will be calibrated during the same fitting process that calibrates the parameters of the neural network. The following equation shows how the first layer of a neural network incorporating embeddings might be written:

\begin{equation}
  z^1 = \sigma(a_1.X'+ b_1), 
\label{eq:embed_to_nn}
\end{equation}

where we represent the output of embedding layers concatenated together with any numerical inputs as \(X'\).

\hypertarget{attention}{%
\subsubsection{Attention}\label{attention}}

Attention mechanisms have been widely applied within the deep learning literature to give more flexibility to RNNs (Bahdanau, Cho, and Bengio 2015) and subsequently as a replacement for FCNs in the so-called Transformer architecture, which is now widely used in natural language processing (Vaswani et al. 2017b), and, more recently, has been applied to computer vision and tabular modeling tasks. Attention mechanisms allow deep neural nets to weight the covariates entering the model in a flexible manner; these types of applications in natural language processing are demonstrated in, for example, Vaswani et al. (2017b). In an actuarial context, attention mechanisms can be understood as giving greater weight to covariates that are important for the modeling task. In this section we provide a theoretical introduction to attention mechanisms and then discuss applications of attention to modeling tabular data. In the following sections of this paper, we then consider whether adding attention to an model that uses embeddings can provide better predictive accuracy. The following is intended as a high level review of some of the relevant theory behind attention based approaches; for more technical details the sources quoted in this section can be consulted.

At a high level, attention mechanisms reweight the covariates \(X\) within a predictive model to give greater weight to covariates that are more predictive for a particular problem. The attention weights can be derived in several different manners. Earlier works, such as Bahdanau, Cho, and Bengio (2015) used so-called ``additive'' attention scores derived as the output of a neural network taking the covariates as inputs. Currently, the most popular form of attention is the ``scaled dot-product'' attention of Vaswani et al. (2017b) which we describe in what follows. We also mention that usually, so-called ``self-attention'' is applied, meaning to say that the importance scores for the covariates are derived using the covariates themselves (i.e.~in what follows, the attention scores are derived in Equation \ref{eq:attention_scores} using the transformed covariates). Other options, such as using a different source of data to derive the attention scores, are also possible.

An attention mechanism is a mapping:

\begin{equation}
X^*: \mathbb{R}^{(q \times d) \times (q \times d) \times (q \times d)} \to \mathbb{R}^{(q \times d)}, ~~
(Q, K, X') ~\mapsto ~ X^* = Attn(Q, K, X'),
\label{eq:attention}
\end{equation}

where \(Q \in \mathbb{R}^{q \times d}\) is a matrix of query vectors, \(K \in \mathbb{R}^{q \times d}\) is a matrix of key vectors, \(X' \in \mathbb{R}^{q \times d}\) is a matrix of value vectors and \(X^* \in \mathbb{R}^{q \times d}\), the output of the attention mechanism, is a new matrix of values that have been reweighted according to their importance for the modeling problem. As noted above, several different options for the mapping in Equation \ref{eq:attention} exist in the literature; here we describe the scaled dot-product self-attention mechanism used in Vaswani et al. (2017b). In more detail, consider that \(X'\) is a matrix of covariates relating to a regression problem, which, in the case of tabular data, will usually be composed of embeddings of the same dimension for several different categorical or numerical variables. In the case discussed here we have \(q\) covariates mapped to embeddings of dimension \(d\). For clarity, note that in Equation \ref{eq:embed_to_nn} we have used the symbol \(X'\) to denote the vector of embeddings and numerical inputs which is passed into the neural network. We slightly overload this notation and, in what follows, we now use the symbol \(X'\) to refer to the matrix of embeddings of dimension \(q \times d\).

We wish to apply the attention mechanism to \(X'\) to assign weights to the most important embeddings for the regression problem; this is done from the ``perspective'' of each covariate represented by the \(q\) rows of \(X'\). To do this, we formulate a so-called query matrix \(Q\) which contains information about the regression problem at hand; this is usually done by passing \(X'\) through a neural network trained to derive queries from the covariates. Similarly, the matrix \(K\) contains the relevance of each row of \(X'\) for the regression problem and is also usually derived by passing each row of \(X'\) through another neural network. Finally, the matrix \(X'\) can be left as is, or processed through another neural network. The first step of the attention mechanism is to derive a matrix \(A \in \mathbb{R}^{q \times q}\) of attention scores:

\begin{equation}
A: \mathbb{R}^{(q \times d) \times (q \times d)} \to \mathbb{R}^{q \times q}, ~~
(Q, K) ~\mapsto ~ A = \frac{Q~K^T}{\sqrt{d}},
\label{eq:attention_scores}
\end{equation}

and where the division by \(\sqrt{d}\) is done element-wise. The attention scores \(A\) are then normalized, or made to sum to unity, across the columns of \(A\) using a so-called softmax function i.e.

\begin{equation}
A^*_{i,j} = \frac{e^{(A_{i,j})}}{\sum_{n=1}^{q} e^{(A_{i,n})}}.
\label{eq:softmax}
\end{equation}

Finally, a linear combination of \(X'\) is formed to give the new matrix of covariates \(X^*\):

\begin{equation}
X^*: \mathbb{R}^{(q \times q) \times (q \times d)} \to \mathbb{R}^{q \times d}, ~~
(A^*, X') ~\mapsto ~ X^* = A^*~X'.
\label{eq:attention_output}
\end{equation}

An intuitive explanation of the attention formula is as follows: we start by considering the importance of all of the covariates that we have for our modeling problem in the context of each individual covariate. This is done by assigning a score between zero and one to each covariate in the modeling problem, and we repeat this process for each of the covariates. Then, we replace each of the original covariates with a new covariate that has been formed by taking a weighted average of the covariates and the scores.

\hypertarget{transformer-models}{%
\subsubsection{Transformer models}\label{transformer-models}}

Self-attention is the main building block of the current state of the art model for natural language processing, which is the Transformer model (Vaswani et al. 2017b). A basic Transformer model applies self-attention to the covariate matrix, \(X'\) which is fed into the model, to produce a new matrix of covariates, \(X^*\). These two matrices, the input matrix \(X'\) and the output matrix \(X^*\) are added (in an element-wise manner) together and then normalized. This matrix is then further processed through a neural network and subsequently, these processed covariates are used in the machine learning task. These stages of the Transformer model help to make the optimization of the network easier than applying scaled self-attention directly to the matrix of covariates \(X'\).

Multiple Transformer network ``blocks'' can be added to a network to create a deep Transformer network.

The Transformer model is much more accurate on NLP tasks compared to using the raw covariates before processing with a Transformer, to the extent that this model now underlies most NLP applications that use deep learning.

\hypertarget{attention-modeling-for-tabular-data}{%
\subsubsection{Attention modeling for Tabular Data}\label{attention-modeling-for-tabular-data}}

Several different approaches to applying attention within the modeling of tabular data have been proposed in the literature. Perhaps the simplest way of doing this is to insert an attention layer between the embeddings and the rest of the neural network. Mathematically, we can modify Equation \ref{eq:embed_to_nn} as follows:

\begin{equation}
\begin{split}
  X^* &= A(X') \\
  z^1 &= \sigma(a_1.X^* + b_1),
\end{split}
\label{eq:embed_to_attn_to_nn}
\end{equation}

where \(X' \in \mathbb{R}^{q \times d}\) is the matrix of embeddings after concatenating \(q\) embeddings of dimension \(d\) together, \(X^*\) represents the matrix of embeddings \(X'\) after processing with attention and \(A\) is the attention function defined in Equations \ref{eq:attention} - \ref{eq:attention_output}. Below, we demonstrate the effect of adding this simple application of attention to a tabular model.

Huang et al. (2020) refines this approach in a model they call the TabTransformer, which implements two main changes to the approach shown in Equations \ref{eq:embed_to_attn_to_nn}. The first of these is that, instead of using the matrix of embeddings after processing with self-attention (\(X'\)) directly in the model, the output of a Transformer model is used instead. The second of these refinements is that for each of the \(q\) covariates with an embedding in the matrix \(X'\), we include a new row embedding which denotes which covariate each embedding relates to. Thus, in the TabTransformer model, we augment the embedding matrix with extra columns, so that the dimension of this matrix is \(q \times (d + d_{col})\), where \(d_{col}\) is the dimension of the row embedding. Huang et al. (2020) find that including this identifier improves the performance of the model. We compare the results of the TabTransformer model to the other models tested below.

Finally, a different way of incorporating attention into the modeling of tabular data has been proposed by Arık and Pfister (2019), who refer to their model as TabNet. In summary, the main idea of the TabNet model is to try emulate the excellent performance of decision trees on tabular data modeling using deep neural networks. This is done by structuring the neural networks to emulate a key aspect of decision tree modeling, which is the efficient selection of relevant covariates for the model task. Within TabNet, an attention based model is used to select features by estimating attention scores for each covariate within \(X\). The covariates are multiplied by the attention scores to down-weight less relevant features for the prediction problem. Other less important details of the TabNet model are given in Arık and Pfister (2019).

\hypertarget{modeling}{%
\section{Predictive Modeling with Embeddings}\label{modeling}}

In this section, we walk through a simple predictive modeling exercise in order to illustrate the application of embeddings. First, we pose a supervised regression problem, based on flood insurance claims severity. We then develop GLMs and neural networks utilizing embeddings for this problem, and inspect the trained parameters. Finally, we provide details of model optimization, inference, and performance.

\hypertarget{problem-description-and-experiment-setup}{%
\subsection{Problem Description and Experiment Setup}\label{problem-description-and-experiment-setup}}

The working example for our experiments is as follows: Given a set of claim characteristics, we predict the losses paid on the property coverage of the policy. The data we use comes from the NFIP and is made available by the OpenFEMA initiative of the Federal Emergency Management Agency (FEMA)\footnote{\url{https://www.fema.gov/about/reports-and-data/openfema}}. Two datasets are made available by OpenFEMA: A policies dataset with exposure information, and a claims dataset with claims transactions, including paid amounts. Because there is no way to associate records of the two datasets, we are limited to fitting severity models on the claims dataset. While the complete dataset contains over two million transactions, for the purposes of our experiments, we limit ourselves to data from 2000 to 2019 and further subsample 100K data points to make experiments feasible on a CPU. The dataset includes a rich variety of variables, from occupancy type to flood zone, and a brief exploratory data analysis can be found in Appendix \ref{EDA}. For our models, we work with a few selected variables that represent continuous and discrete variables of low and high cardinalities, which we list in Table \ref{tab:variables}

\begin{table}[!h]

\caption{\label{tab:variables}Predictor variables.}
\centering
\begin{tabular}[t]{ll}
\toprule
Variable & Type\\
\midrule
Building insurance coverage & Numeric\\
Community Rating System Discount & Numeric\\
Basement enclosure type & Categorical\\
Occupancy Type & Categorical\\
Number of floors in the insured building & Categorical (binned in original dataset)\\
Flood zone & Categorical (high cardinality)\\
Primary residence & Categorical (binary indicator)\\
\bottomrule
\end{tabular}
\end{table}

The response variable we take from the dataset is ``Amount Paid on Building Claim,'' which is a numeric variable.

For our experiments, we set up a 5-fold cross validation scheme that is applied to each of the models we introduce. In Sections \ref{models} and \ref{interpretation}, the tables and figures are based on a single fold of the cross validation split. In Section \ref{training-inference}, we provide the cross-validated performance metrics of each of the models.

We note that the focus of our contribution is on applications of categorical embeddings in predictive modeling, rather than on fine-tuning models in order to beat or establish a benchmark. To this end, we do not perform systematic hyperparameter tuning in our experiments, and the performance metrics are provided to show that reasonable models utilizing categorical embeddings can be developed. However, as we describe the modeling procedure and results, we provide commentary on recommended practices in practical applications, especially with respect to the neural network models.

\hypertarget{models}{%
\subsection{Models}\label{models}}

In this subsection, we develop the following models:

\begin{enumerate}
\def\labelenumi{\arabic{enumi}.}
\tightlist
\item
  A GLM with gamma distribution and log link function,
\item
  A neural network with one-dimensional categorical embeddings,
\item
  A GLM with the categorical predictors replaced by the learned embeddings from Model 2,
\item
  A neural network with multidimensional categorical embeddings.
\end{enumerate}

These model architectures are relatively uncomplicated by today's standards, and are so chosen to better highlight the embedding components. Later, in Section \ref{attention_models}, we will investigate more involved architectures utilizing embeddings that represent the state of the art for modeling tabular data.

\hypertarget{model-1-glm}{%
\subsubsection{Model 1: GLM}\label{model-1-glm}}

While this paper focuses on embeddings in neural networks, we begin with a GLM to provide a common frame of reference, since most actuaries are familiar with the technique. Although GLMs are commonplace and well-studied in the actuarial literature, there are still plenty of decisions to be made in the modeling process; see, for example, Goldburd et al. (2020) for an in-depth discussion. For our purpose of establishing a baseline, we proceed with what we perceive as reasonable decisions around feature engineering and model structure, outlined below:

\begin{itemize}
\item
  Target variable: Amount paid on building claim
\item
  Predictors:

  \begin{itemize}
  \tightlist
  \item
    \(\log\) \texttt{building\_insurance\_coverage}
  \item
    \texttt{basement\_enclosure\_type},
  \item
    \texttt{number\_of\_floors\_in\_the\_insured\_building}
  \item
    \emph{prefix} of \texttt{flood\_zone}
  \item
    \texttt{primary\_residence}
  \end{itemize}
\item
  Link function: \(\log\)
\item
  Distribution: gamma
\end{itemize}

We take the log of the continuous predictor \texttt{building\_insurance\_coverage}, which allows the scale of the predictor to match that of the target variable. Because the \texttt{flood\_zone} variable in the original data contains 60 levels, we take the prefix of the zone code, which corresponds to the level of risk as determined by FEMA.\footnote{\url{https://www.floodsmart.gov/flood-map-zone/find-yours}} For example, \texttt{"A01"}, \texttt{"A02"}, and so on are recoded as simply \texttt{"A"}. A log link together with the gamma distribution is a standard choice for severity modeling, which provides a multiplicative structure where the response is positive. Following the notation in \ref{catdata}, the number of parameters in the GLM is

\begin{equation}
1 + |J^{\text{num}}| + \left(\sum_{j \in J^{\text{cat}}} n_{\mathcal{P}}^j\right) - |J^{\text{cat}}|:
\label{eq:glm1}
\end{equation}

one for the intercept, one for each numeric variable, and the numbers of levels minus one (due to dummy coding) for each categorical variable.

In Table \ref{tab:relativities}, we exhibit an exerpt of the relativities, or exponentiated fitted coefficients, of a couple of the categorical variables. Recall that the choice of the log link gives a multiplicative structure, so the interpretation of 1.34 for \texttt{occupancy\_type\ =\ "Non-residential\ building"} is that, all else equal, the expected loss for a policy for a non-residential building is 1.34 times that of a policy for a single family home, which is the base, or reference, level for the factor.

\begin{longtable}[]{@{}llr@{}}
\caption{\label{tab:relativities} Excerpt of relativities for Model 1.}\tabularnewline
\toprule
variable & label & relativity\tabularnewline
\midrule
\endfirsthead
\toprule
variable & label & relativity\tabularnewline
\midrule
\endhead
occupancy\_type & Single family residence & 1.00\tabularnewline
occupancy\_type & 2 to 4 unit residential buidling & 1.10\tabularnewline
occupancy\_type & Non-residential building & 1.34\tabularnewline
occupancy\_type & Residential building with more than 4 units & 1.59\tabularnewline
flood\_zone & C & 1.00\tabularnewline
flood\_zone & A & 1.42\tabularnewline
flood\_zone & X & 1.18\tabularnewline
flood\_zone & B & 1.65\tabularnewline
flood\_zone & V & 1.38\tabularnewline
flood\_zone & D & 1.45\tabularnewline
\bottomrule
\end{longtable}

\hypertarget{model-2-neural-network-with-unidimensional-embeddings}{%
\subsubsection{Model 2: Neural Network with Unidimensional Embeddings}\label{model-2-neural-network-with-unidimensional-embeddings}}

\begin{figure}

{\centering \includegraphics[width=0.8\linewidth,]{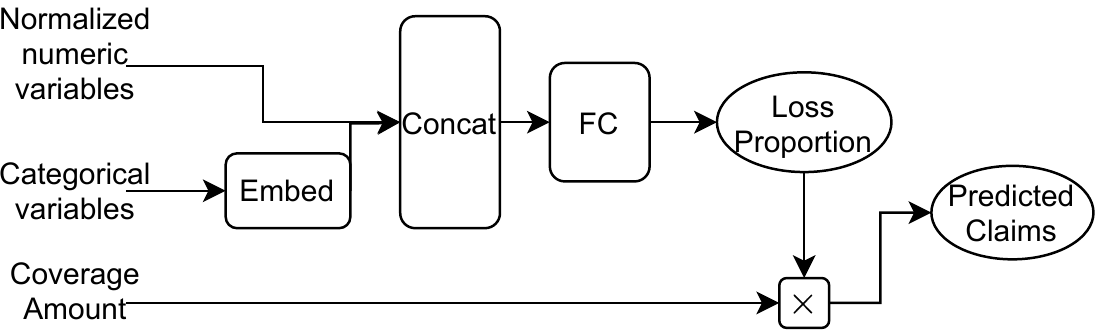} 

}

\caption{Architecture of Models 2 and 4.}\label{fig:nn-1}
\end{figure}

For Model 2, we build a feedforward neural network, whose architecture is shown in Figure \ref{fig:nn-1}. The categorical inputs go through one-dimensional embedding layers where they are each mapped to a scalar. The embeddings are then concatenated with the numeric predictors, which have been normalized in data pre-processing, before being passed through a feedforward layer (with 8 hidden units and ReLU activation) to obtain a scalar value between 0 and 1, as constrained by a sigmoid output activation. This scalar output represents the proportion of coverage amount that is paid out, which we then multiply by the dollar amount of the coverage amount to obtain our prediction of the claims paid. For optimization, we use the mean squared error (MSE) loss function.

With a single hidden layer, this architecture is relatively simple; without the hidden layer and corresponding ReLU activation, we would actually recover a type of logistic regression structure. The choice of one for the embedding dimension is also due to simplicity: it is easily interpretable as the representation of each factor level as a point on the real number line; also, the trained embeddings are easily incorporated into a GLM which we will see in Model 3. In practice, the embedding dimension is a hyperparameter one can tune, for example via cross validation, and can vary for each variable. However, we also note that the choice of one for the embedding output dimension is not an unreasonable one, especially when the cardinalities of the categorical variables are not large (i.e., when they are in the 10s).

In Tables \ref{tab:embeddings-occupancy-type} and \ref{tab:embeddings-flood-zone}, we show the learned embeddings for \texttt{occupancy\_type} and an excerpt of them for \texttt{flood\_zone}. While it may be tempting to compare these values to those in Table \ref{tab:relativities}, they represent fundamentally different concepts. It is helpful to think of the embedding values for each variable as a continuous represention of the variable itself, rather than a ``coefficient'' associated with the variable. Note that it is not straightforward to inspect the embeddings from a table and infer relationships to the response variable, even in a shallow network such as ours, due to nonlinearities within the model. Also note that we now have a way to incorporate the full granularity of the \texttt{flood\_zone} variable, which is why more levels are shown in Table \ref{tab:embeddings-flood-zone} as compared to Table \ref{tab:relativities}. In Section \ref{interpretation}, we will discuss techniques to visualize these embeddings.

\begin{longtable}[]{@{}lr@{}}
\caption{\label{tab:embeddings-occupancy-type} Learned embeddings for \texttt{occupancy\_type} in Model 2.}\tabularnewline
\toprule
level & embedding\tabularnewline
\midrule
\endfirsthead
\toprule
level & embedding\tabularnewline
\midrule
\endhead
2 to 4 unit residential buidling & 1.46\tabularnewline
Non-residential building & -1.04\tabularnewline
Residential building with more than 4 units & 0.12\tabularnewline
Single family residence & 0.36\tabularnewline
\bottomrule
\end{longtable}

\newpage

\begin{longtable}[]{@{}lrlrlrlrlr@{}}
\caption{\label{tab:embeddings-flood-zone} Excerpt of learned embeddings for \texttt{flood\_zone} in Model 2.}\tabularnewline
\toprule
level & embedding & level & embedding & level & embedding & level & embedding & level & embedding\tabularnewline
\midrule
\endfirsthead
\toprule
level & embedding & level & embedding & level & embedding & level & embedding & level & embedding\tabularnewline
\midrule
\endhead
A & 1.48 & A09 & -0.48 & A18 & 0.39 & A28 & 1.19 & C & 1.73\tabularnewline
A00 & -1.05 & A0B & 1.47 & A19 & -1.06 & A30 & -0.01 & D & -1.15\tabularnewline
A01 & -0.18 & A10 & -0.25 & A20 & -1.01 & A99 & -0.44 & V & -1.85\tabularnewline
A02 & -0.06 & A11 & 0.35 & A21 & 2.08 & AE & 1.12 & V01 & 0.02\tabularnewline
A03 & -0.01 & A12 & -0.38 & A22 & -0.15 & AH & 2.19 & V02 & -1.55\tabularnewline
A04 & 0.05 & A13 & 2.81 & A23 & -0.76 & AHB & -0.68 & V03 & -0.98\tabularnewline
A05 & 1.30 & A14 & 0.58 & A24 & -0.87 & AO & 0.55 & V04 & 1.73\tabularnewline
A06 & -0.76 & A15 & -0.91 & A25 & 0.02 & AOB & 0.08 & V05 & -1.97\tabularnewline
A07 & -0.62 & A16 & 0.79 & A26 & 0.00 & AR & 2.93 & V06 & 0.43\tabularnewline
A08 & -0.53 & A17 & 1.35 & A27 & 0.80 & B & 0.33 & V07 & 1.91\tabularnewline
\bottomrule
\end{longtable}

\hypertarget{model-3-glm-with-neural-network-embeddings}{%
\subsubsection{Model 3: GLM with Neural Network Embeddings}\label{model-3-glm-with-neural-network-embeddings}}

For Model 3, we return to the GLM, but now we replace the categorical variables with the trained embeddings from Model 2. In other words, our model now contains only continuous variables, and each categorical factor is represented as a scalar value and obtains its own coefficient. In contrast with the number of parameters computed in Equation \eqref{eq:glm1}, we have fewer parameters for this model:

\begin{equation}
1 + |J^{\text{num}}| + |J^{\text{cat}}|.
\label{eq:glm2}
\end{equation}
Hence, the model can be written as

\begin{equation}
\mu_i = \beta_0 + \sum_{j \in J^{\text{num}}}\beta_{j} x_{i,j} + \sum_{j \in J^{\text{cat}}} \beta_{j} z_{i,j},
\end{equation}

where \(z_{i,j}\) represents the embedding value for \(i\)th instance of the categorical column \(X_j\).

Although, in this particular example, we incorporate one-dimensional embeddings, one can easily include multidimensional embeddings, which would increase the model's flexibility. The modeler would need to balance the potential additional lift with overfitting and some loss of interpretability, since, with one-dimensional embeddings, our GLM structure dictates that there is a monotonic relationship between the predicted response and the input embedding, which could be desirable in certain use cases (and undesirable in others.)

Since we only have a few parameters in this model, we exhibit the full list of fitted coefficients in Table \ref{tab:glm2}.

\begin{longtable}[]{@{}lr@{}}
\caption{\label{tab:glm2} Fitted coefficients of Model 3.}\tabularnewline
\toprule
term & estimate\tabularnewline
\midrule
\endfirsthead
\toprule
term & estimate\tabularnewline
\midrule
\endhead
(Intercept) & 3.371\tabularnewline
total\_building\_insurance\_coverage & 0.586\tabularnewline
basement\_enclosure\_crawlspace\_type & 0.597\tabularnewline
number\_of\_floors\_in\_the\_insured\_building & -0.278\tabularnewline
occupancy\_type & -0.075\tabularnewline
flood\_zone & -0.007\tabularnewline
primary\_residence & -0.062\tabularnewline
community\_rating\_system\_discount & 0.706\tabularnewline
\bottomrule
\end{longtable}

\hypertarget{model-4-neural-network-with-multidimensional-embeddings}{%
\subsubsection{Model 4: Neural Network with Multidimensional Embeddings}\label{model-4-neural-network-with-multidimensional-embeddings}}

Model 4 is architecturally identical to Model 2, with the only difference being the number of embedding dimensions: instead of mapping factor levels to real numbers, we map them to points in Euclidean space, where the dimension of the space is the ceiling of the factor's cardinality divided by two. Formally, in the notation of Section \ref{catdata}, \(q_{\mathcal{P}^j}=\ceil{n^j_{\mathcal{P}}/2}\) for variable \(X_j\). Following the discussion thus far, we list the learned embeddings for a couple variables in Tables \ref{tab:nn2-occupancy-type} and \ref{tab:nn2-flood-zone}. Note that, for \texttt{flood\_zone}, we are only displaying the embedding for \emph{one} level.

\begin{longtable}[]{@{}lrrr@{}}
\caption{\label{tab:nn2-occupancy-type} Learned embeddings for \texttt{occupancy\_type} in Model 4.}\tabularnewline
\toprule
level & e1 & e2 & e3\tabularnewline
\midrule
\endfirsthead
\toprule
level & e1 & e2 & e3\tabularnewline
\midrule
\endhead
2 to 4 unit residential buidling & 0.37 & -1.76 & 1.36\tabularnewline
Non-residential building & 0.13 & 1.47 & 1.77\tabularnewline
Residential building with more than 4 units & -1.12 & -0.15 & -1.43\tabularnewline
Single family residence & 0.85 & 0.16 & -0.99\tabularnewline
\bottomrule
\end{longtable}

\begin{longtable}[]{@{}llllllll@{}}
\caption{\label{tab:nn2-flood-zone} Learned embeddings for level \texttt{"A00"} of variable \texttt{flood\_zone} in Model 4.}\tabularnewline
\toprule
dimension & value & dimension & value & dimension & value & dimension & value\tabularnewline
\midrule
\endfirsthead
\toprule
dimension & value & dimension & value & dimension & value & dimension & value\tabularnewline
\midrule
\endhead
e1 & -0.167 & e11 & -0.810 & e21 & -0.300 & e31 & -0.421\tabularnewline
e2 & -0.363 & e12 & 1.380 & e22 & -0.194 & e32 & 0.594\tabularnewline
e3 & -0.893 & e13 & -0.180 & e23 & -0.320 & e33 & 1.681\tabularnewline
e4 & 0.130 & e14 & -1.612 & e24 & -0.035 & e34 & 1.668\tabularnewline
e5 & -1.588 & e15 & -1.327 & e25 & -1.128 & &\tabularnewline
e6 & -0.123 & e16 & -0.965 & e26 & 0.168 & &\tabularnewline
e7 & -0.259 & e17 & -1.356 & e27 & -1.022 & &\tabularnewline
e8 & 0.324 & e18 & -0.061 & e28 & 1.011 & &\tabularnewline
e9 & -1.430 & e19 & -0.120 & e29 & -0.843 & &\tabularnewline
e10 & -1.598 & e20 & -0.488 & e30 & -0.743 & &\tabularnewline
\bottomrule
\end{longtable}

\hypertarget{interpretation}{%
\subsection{Visualization Techniques for Interpretation}\label{interpretation}}

With AI and ML coming into the mainstream across industries and use cases, the topic of model explainability has received increasing attention. In insurance pricing, Kuo and Lupton (2020) propose a framework for model interpretability in the context of regulation and actuarial standards of practice, while Henckaerts et al. (2020) investigate algorithm-specific techniques to inspect tree-based models. Kuo and Lupton (2020) and references therein also discuss commonly utilized model-agnostic explanation techniques, such as permutation variable importance and partial dependence plots, that can be applied to the neural networks we consider in this paper.

In this section, we focus on ways to visualize the learned embeddings. In the authors' experiences, oftentimes the formats of the visualizations---basic line and scatter plots---even more so than the content of the data, are helpful in explaining the concept to stakeholders.

\begin{figure}

{\centering \includegraphics[width=0.8\linewidth,]{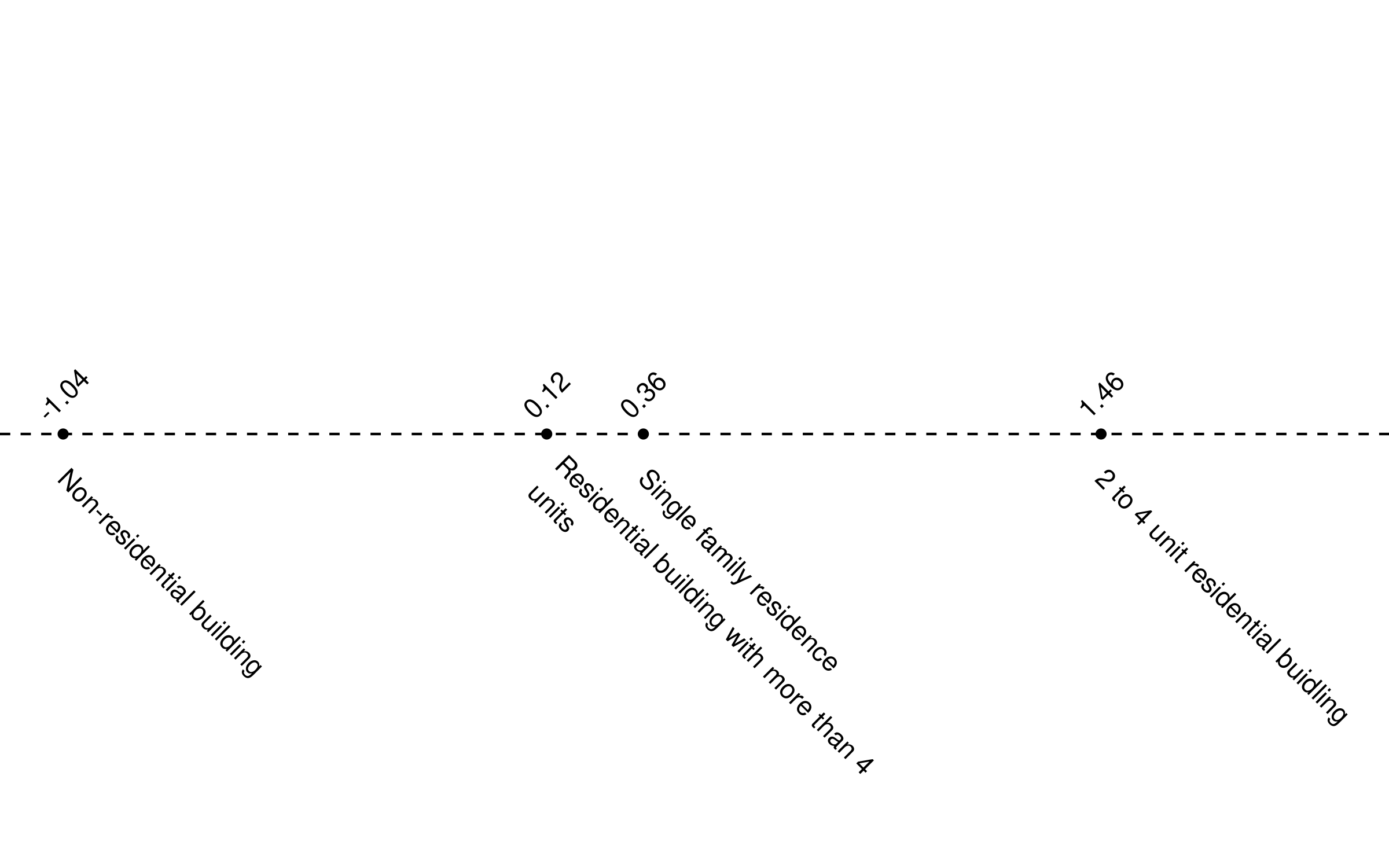} 

}

\caption{Learned embeddings of occupancy type visualized on a number line.}\label{fig:occ-type}
\end{figure}

\begin{figure}

{\centering \includegraphics[width=0.6\linewidth,]{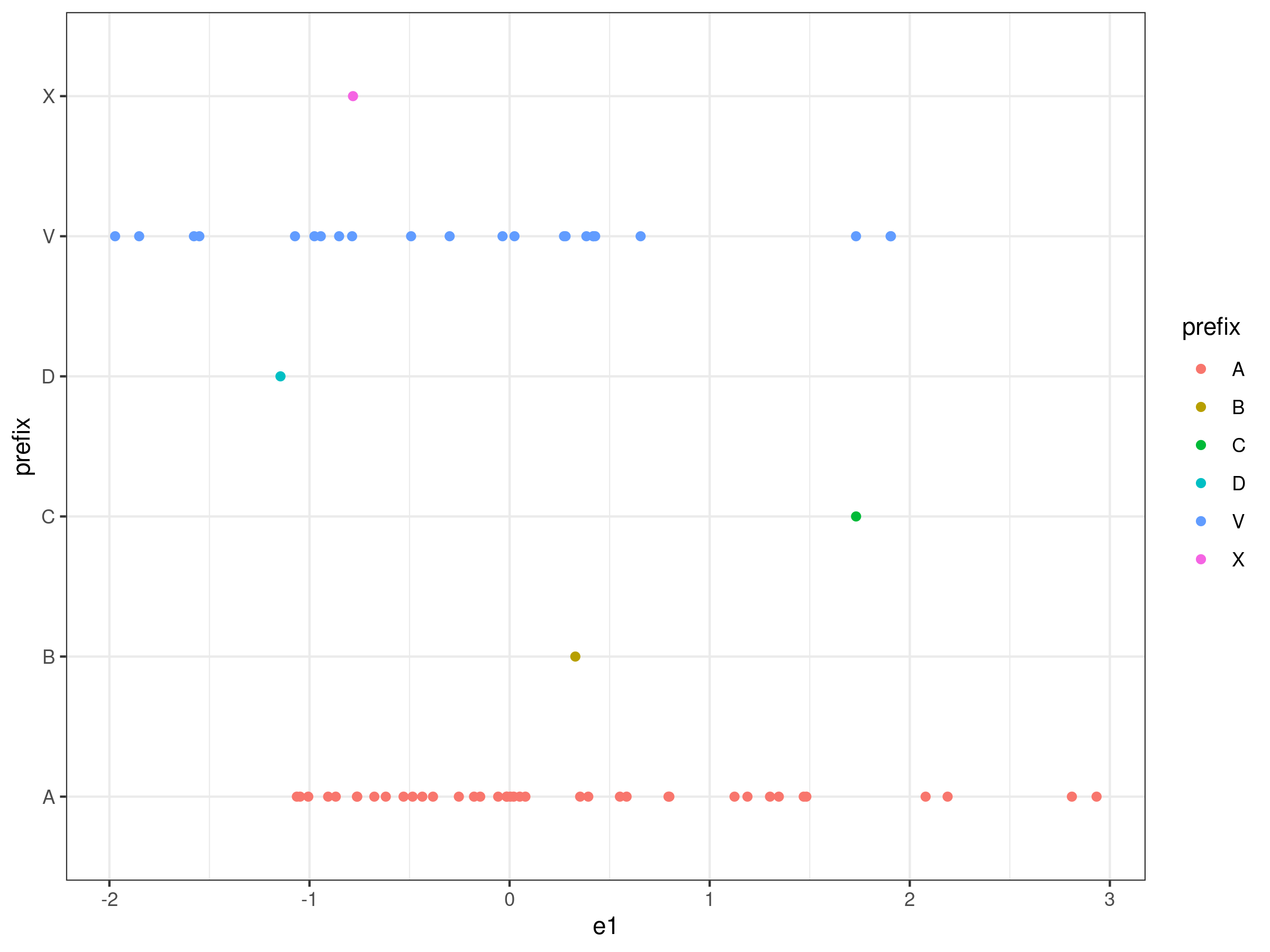} 

}

\caption{Learned embeddings of flood zone by prefix.}\label{fig:flood-zone}
\end{figure}

For Model 2, the neural network with one-dimensional embeddings, the task is simple: we can simply plot the levels on a line, as in Figure \ref{fig:occ-type}. In the case of \texttt{flood\_zone}, where we happen to have the luxury of knowing a priori the meaning of the prefixes, we can plot them separately, as in Figure \ref{fig:flood-zone} to see how the model mapped them.

For Model 4, the neural network with multi-dimensional embeddings, we would need to treat the learned values before we can plot them. Specifically, we need to reduce the dimension of the data to 2D for embeddings with more than two dimensions. The standard approachs for this are principal component analysis (PCA; Shlens (2014)) and t-distributed stochastic neighbor embedding (t-SNE; Maaten and Hinton (2008)), both of which we illustrate.

\hypertarget{principal-component-analysis}{%
\subsubsection{Principal Component Analysis}\label{principal-component-analysis}}

In both approaches, the overarching theme is as follows: we have a collection of points in higher dimensional space, \(\mathcal{X} = \{x_1, x_2, \dots, x_n\}\) that we wish to map to some lower dimensional space, typically \(\mathbb{R}^2\), obtaining \(\mathcal{Y} = \{y_1, y_2, \dots, y_n\}\). PCA performs a linear change of coordinates on the input data so that the variance of the data is maximized along each orthogonal coordinate, also known as \emph{principal component}, successively. Formally, if we denote \(\mathbf{X}\) and \(\mathbf{Y}\) to be the matrices whose columns are elements of \(\mathcal{X}\) and \(\mathcal{Y}\), respectively, this can be represented as looking for an orthonormal matrix \(\mathbf{P}\) in \(\mathbf{Y} = \mathbf{PX}\) such that the covariance matrix of \(\mathbf{Y}\), \(\frac{1}{n}\mathbf{YY^T}\), is diagonal. Here, the rows of \(\mathbf{P}\) are the principal components that define the resulting coordinate system. It turns out that that these principal components are the eigenvectors of the covariance matrix of \(\mathbf{X}\), i.e., \(\mathbf{C_X} \equiv \frac{1}{n}\mathbf{XX^T}\). In practice, numerical procedures, including R's \texttt{prcomp()} which we use, utilize singular value decomposition (SVD) to compute the eigenvectors.

\begin{figure}

{\centering \includegraphics[width=0.5\linewidth,]{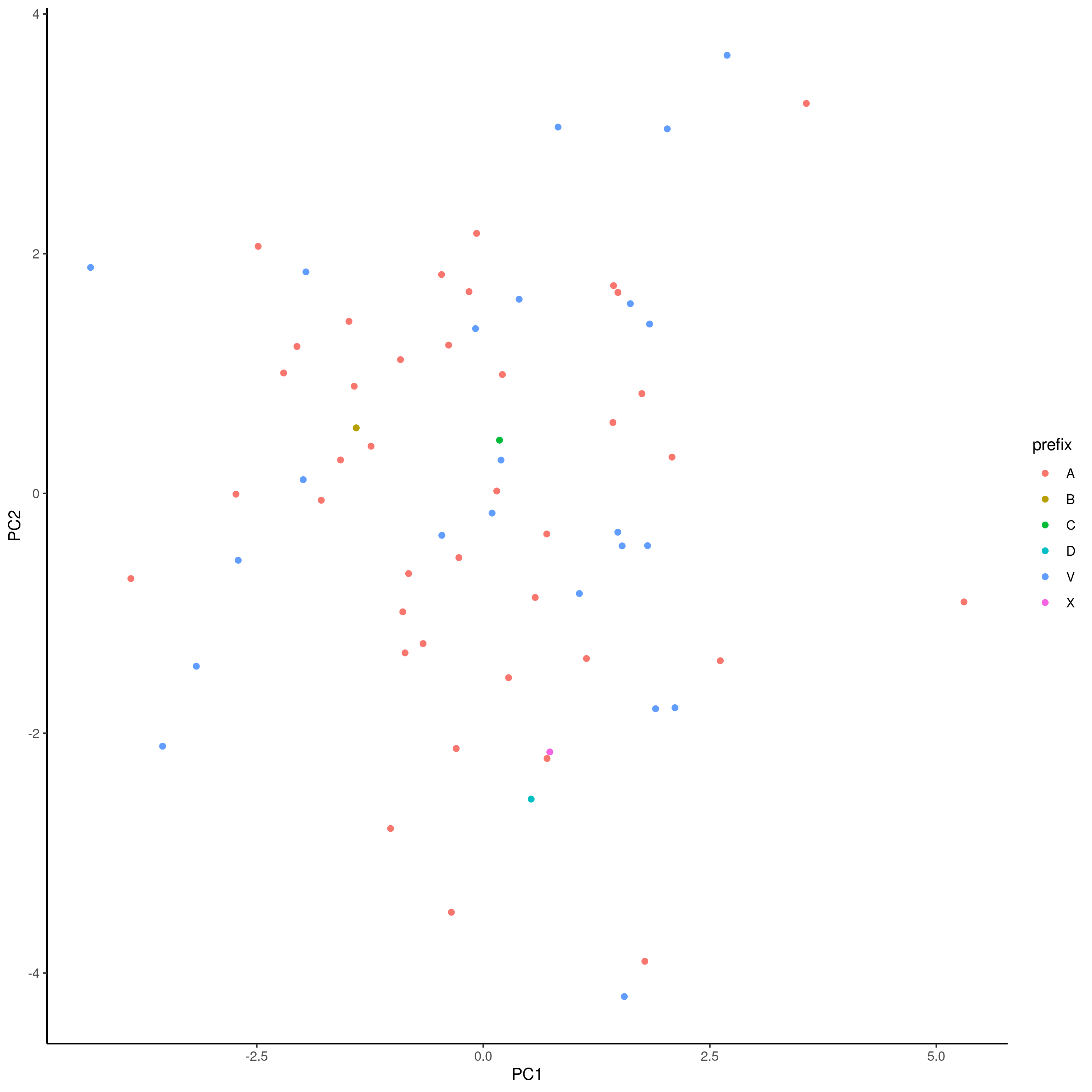} 

}

\caption{First two principal components of learned embeddings of flood zone in Model 4.}\label{fig:flood-zone-pca}
\end{figure}

In Figure \ref{fig:flood-zone-pca}, we plot the learned embeddings for \texttt{flood\_zone} in Model 3 on the coordinate system defined by the first two principle components. In our case, however, we are not able to identify discernable clusters in the plot. From inspecting the matrix decomposition results, we see that the cumulative proportion of variance for the first two principal components are just 0.18, which may suggest that perhaps PCA is not the most appropriate dimensionality reduction technique in this case.

\hypertarget{t-distributed-stochastic-neighbor-embedding}{%
\subsubsection{t-Distributed Stochastic Neighbor Embedding}\label{t-distributed-stochastic-neighbor-embedding}}

In contrast with PCA, t-SNE is a nonlinear technique. At a high level, it computes a similarity metric, in the form of conditional probabilities based on pairwise distances, among points in \(\mathcal{X}\), then look for points \(\mathcal{Y}\) in the lower dimensional space which have similarity metric values (for a differently defined metric) that are ``close'' to those of corresponding pairs of points in the original data.

Let \(p_{j|i}\) denote the probability that point \(x_i\) picks point \(x_j\) as its \emph{neighbor}, if it picks neighbors in proportion to its density under an \(n\)-dimension Gaussian centered at \(x_i\). Formally, we can write \(p_{j|i}\) as

\begin{equation}
p_{j|i} = \frac{\exp(\| x_i - x_j \|^2 / 2\sigma_i^2)}{\sum_{k\neq i} \exp(-\| x_i - x_k \|^2 / 2\sigma_i^2)}, \label{eq:pji}
\end{equation} where \(\sigma_i^2\) denotes the variance of the Gaussian associated with \(x_i\). The similarity metric for the original space symmetrizes this conditional probability and is defined as

\begin{equation}
p_{ij} = \frac{p_{j|i} + p_{i|j}}{2n}.
\end{equation}

On the other hand, in the target space, t-SNE assumes the neighbor selection is driven by a Student t-distribution with one degree of freedom, i.e., a Cauchy distribution. The similarity metric for a pair of points in \(\mathcal{Y}\) is then defined as the joint probability

\begin{equation}
q_{ij} = \frac{(1 + \| y_i - y_j \|^2)^{-1}}{\sum_{k \neq l} (1 + \| y_k - y_l \|^2)^{-1}}.
\end{equation}

With the similarity metrics defined for each space, we can then state the objective for ``closeness'' for the sets of points, which is the Kullback--Leibler divergence from Q to P:

\begin{equation}
D_{KL}(P || Q) = \sum_{i \neq j} p_{ij} \log \frac{p_{ij}}{q_{ij}}. \label{eq:tsne-obj}
\end{equation}

We now circle back to Equation \eqref{eq:pji} and discuss \(\sigma_i\). The variance of the Gaussian is allowed to vary depending on where the point is in the space; specifically, it should reflect the density or sparsness of surrounding points. To train this parameter, t-SNE introduces a hyperparameter, \emph{perplexity}, which is a scalar value set by the user, and is defined as

\begin{equation}
Perp(P_i) = 2^{H(P_i)},
\end{equation}
where \(H(P_i)\) denotes the entropy of \(P_i\),

\begin{equation}
H(P_i) = -\sum_j p_{j|i}\log_2 p_{j|i},
\end{equation}

and \(P_i\) is the Gaussian parameterized by \(x_i\) and \(\sigma_i\). During training, t-SNE looks for \(P_i\) that have perpelxity corresponding to the user input. Finally, the objective as stated in Equation \eqref{eq:tsne-obj} is differentiable, and is trained via gradient descent.

\begin{figure}

{\centering \includegraphics[width=0.6\linewidth,]{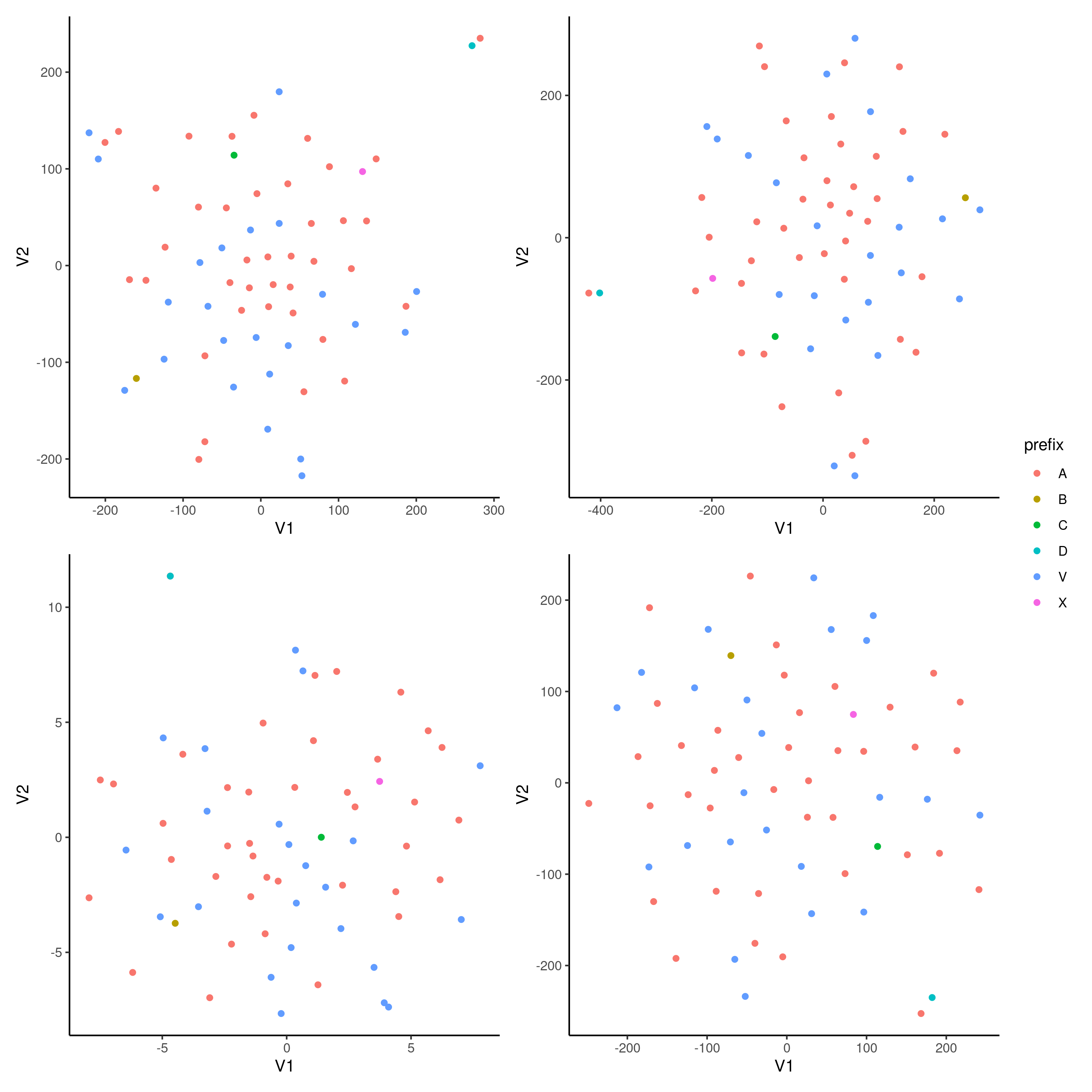} 

}

\caption{t-SNE plots of learned embeddings of flood zone in Model 4, with various perplexity values. From left to right then top to bottom: 2, 3, 5, and 10.}\label{fig:flood-zone-tsne}
\end{figure}

While t-SNE is a powerful visualization technique, there are a few points the modeler should be mindful of in practice. We refer the reader to Wattenberg, Viégas, and Johnson (2016) for a more thorough discussion, but stress here that the resulting plot can be very sensitive to the perplexity hyperparameter. In addition, for our use case, which is to explore the learned embeddings, we have fewer data points than most other applications in the literature. Accordingly, our reasonable range of hyperparameter values need to be adjusted down. In Figures \ref{fig:flood-zone-tsne}, we display the t-SNE of the learned embeddings for \texttt{flood\_zone} in Model 3 with various perplexity settings: 2, 3, 5, and 10; the learning rate and the number of steps are fixed at 100 and 10,000, respectively, for all plots. In the first three plots, we can see the semblance of clusters forming. In particular, flood zone D appears in its own cluster, which implies it is far away from other levels in the embedding space, \emph{as seen by the model}. When the perplexity is set to 10 (and beyond), however, we start to lose the geometric properties.

\hypertarget{training-inference}{%
\subsection{Training and Inference}\label{training-inference}}

For the GLMs, Models 1 and 3, we follow the standard optimization procedure using iteratively reweighted least squares (IWLS) without regularization. For the neural networks, Models 2 and 4, we utilize the \texttt{Adam} optimizer (Kingma and Ba 2014) with a learning rate of 0.01 and a minibatch size of 1000. We split the training set into analysis and assessment sets, containing 80\% and 20\% of the rows, respectively. We train on the analysis set for a maximum of 15 epochs, and stop early if the loss on the assessment set does not improve for 5 epochs.

During inference, or scoring, care must be taken to accommodate the categorical levels that are unseen during training. While we allow for an extra key in the embedding dictionary, the initial weights are not updated during training, which could lead to nonsensical predictions in the validation or test sets. To circumvent this issue, before applying the trained model but after the final optimization epoch, we update the weights for the unseen levels manually. In practice, a common approach is to take an average value of some sort of the weights of the other levels. In our case, we take the median (if there are multiple, the lower of the two), but a mean, mean weighted by frequency of observations, or trimmed mean could be used.

In Table \ref{tab:cv-results}, we exhibit the cross-validated performance metrics of each of the models. We report both the root mean squared error (RMSE) and the mean absolute error (MAE); we note that these are just two of many potential metrics to use, and are so chosen for their interpretability and prevalence in the literature for comparing different types of models. Recall that, between the two metrics, RMSE penalizes large mispredictions than MAE. Also, in the optimization process, GLM with a gamma distribution optimizes the deviance, rather than MSE directly.

We see that the neural networks generally outperform the GLMs; however, what is more interesting is that using learned embeddings in place of categorical levels improves the GLM, at least from the RMSE perspective.

While linear regression (i.e., the specific case of GLM with a gaussian distribution and identity link function) is not commonly used for severity modeling, we add it into the mix as a benchmark, a practice we recommend in order to check reasonableness of other models' results. Since it is possible for a linear regression to output negative values, we cap the predictions below by 0.01 during scoring. Somewhat surprisingly, in the problem we are considering, it performs similarly to the more complex neural network models in the RMSE metric (which it optimizes for, as the unit deviance of the normal distribution correspond to MSE.) On the other hand, the performance of the linear model is quite significantly worse on the MAE metric. Of course, this is not to discredit neural networks or GLMs, as the experiments are performed without extensive hyperparameter tuning or feature engineering.

\begin{longtable}[]{@{}lrr@{}}
\caption{\label{tab:cv-results} Cross validation results.}\tabularnewline
\toprule
\begin{minipage}[b]{0.76\columnwidth}\raggedright
Model\strut
\end{minipage} & \begin{minipage}[b]{0.08\columnwidth}\raggedleft
RMSE\strut
\end{minipage} & \begin{minipage}[b]{0.08\columnwidth}\raggedleft
MAE\strut
\end{minipage}\tabularnewline
\midrule
\endfirsthead
\toprule
\begin{minipage}[b]{0.76\columnwidth}\raggedright
Model\strut
\end{minipage} & \begin{minipage}[b]{0.08\columnwidth}\raggedleft
RMSE\strut
\end{minipage} & \begin{minipage}[b]{0.08\columnwidth}\raggedleft
MAE\strut
\end{minipage}\tabularnewline
\midrule
\endhead
\begin{minipage}[t]{0.76\columnwidth}\raggedright
Model 1: GLM (gamma/log link)\strut
\end{minipage} & \begin{minipage}[t]{0.08\columnwidth}\raggedleft
65,138\strut
\end{minipage} & \begin{minipage}[t]{0.08\columnwidth}\raggedleft
38,367\strut
\end{minipage}\tabularnewline
\begin{minipage}[t]{0.76\columnwidth}\raggedright
Model 2: MLP with 1-dimensional embeddings\strut
\end{minipage} & \begin{minipage}[t]{0.08\columnwidth}\raggedleft
60,979\strut
\end{minipage} & \begin{minipage}[t]{0.08\columnwidth}\raggedleft
37,605\strut
\end{minipage}\tabularnewline
\begin{minipage}[t]{0.76\columnwidth}\raggedright
Model 3: GLM (gamma/log link), categorical variables replaced with embeddings\strut
\end{minipage} & \begin{minipage}[t]{0.08\columnwidth}\raggedleft
63,170\strut
\end{minipage} & \begin{minipage}[t]{0.08\columnwidth}\raggedleft
38,360\strut
\end{minipage}\tabularnewline
\begin{minipage}[t]{0.76\columnwidth}\raggedright
Model 4: MLP with multidimensional embeddings\strut
\end{minipage} & \begin{minipage}[t]{0.08\columnwidth}\raggedleft
60,973\strut
\end{minipage} & \begin{minipage}[t]{0.08\columnwidth}\raggedleft
37,574\strut
\end{minipage}\tabularnewline
\begin{minipage}[t]{0.76\columnwidth}\raggedright
Linear regression, predictions capped below at 0.01\strut
\end{minipage} & \begin{minipage}[t]{0.08\columnwidth}\raggedleft
60,879\strut
\end{minipage} & \begin{minipage}[t]{0.08\columnwidth}\raggedleft
38,431\strut
\end{minipage}\tabularnewline
\bottomrule
\end{longtable}

\hypertarget{attention_models}{%
\section{Attention based modeling}\label{attention_models}}

In this section we investigate whether the predictive performance of the models applied to the NFIP dataset can be improved by applying the attention based models defined in Section \ref{attention}. We discuss the implementation of each of the Simple Attention, TabTransformer and TabNet models in the following sub-section, then discuss how the modifications made to the embeddings through the application of attention can be visualized in the following sub-section. We conclude the section with a discussion of the model results.

\newpage

\hypertarget{models-1}{%
\subsection{Models}\label{models-1}}

In this subsection, we develop the following models:

\begin{enumerate}
\def\labelenumi{\arabic{enumi}.}
\setcounter{enumi}{4}
\tightlist
\item
  A neural network based on the design of Model 4, with an attention component
\item
  Similar to Model 5, a neural network implementing the TabTransformer architecture
\item
  A neural network implementing the TabNet architecture
\end{enumerate}

Compared to the models presented in the previous section, the models presented here are closer to state of the art models that might be used for tabular data modeling problems. To the best of our knowledge, this is the first analysis of these architectures within the actuarial literature.

\hypertarget{model-5-simple-attention-network}{%
\subsubsection{Model 5: Simple Attention Network}\label{model-5-simple-attention-network}}

\begin{figure}

{\centering \includegraphics[width=0.8\linewidth,]{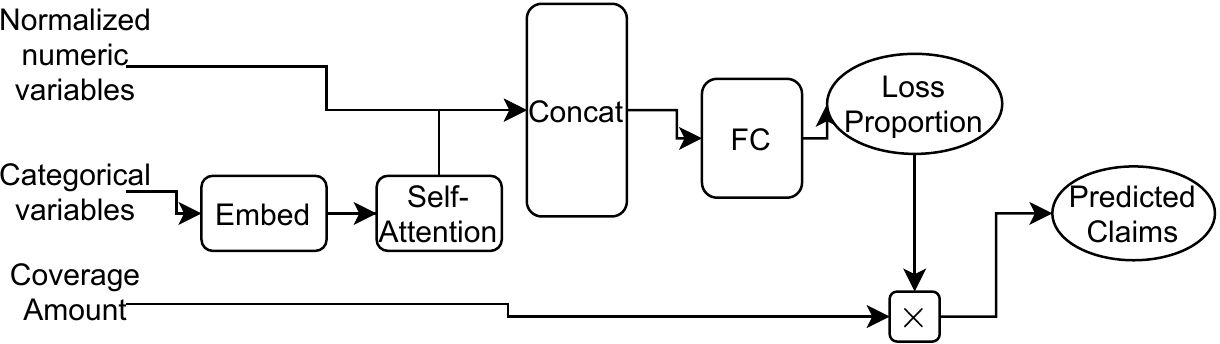} 

}

\caption{Architecture of Model 5.}\label{fig:nn-2}
\end{figure}

Model 5 is conceptually similar to Model 4, except for the following two aspects. First, we standardize the dimension of the embedding layers to 8 dimensions for each categorical covariate. This is done to allow for the application of the attention component, which requires the embeddings for each covariate to have the same dimension. Second, we apply a self-attention mechanism within the network, as shown in Equation \ref{eq:embed_to_attn_to_nn}. We illustrate this architecture in Figure \ref{fig:nn-2}.

\hypertarget{model-6-tabtransformer}{%
\subsubsection{Model 6: TabTransformer}\label{model-6-tabtransformer}}

\begin{figure}

{\centering \includegraphics[width=0.8\linewidth,]{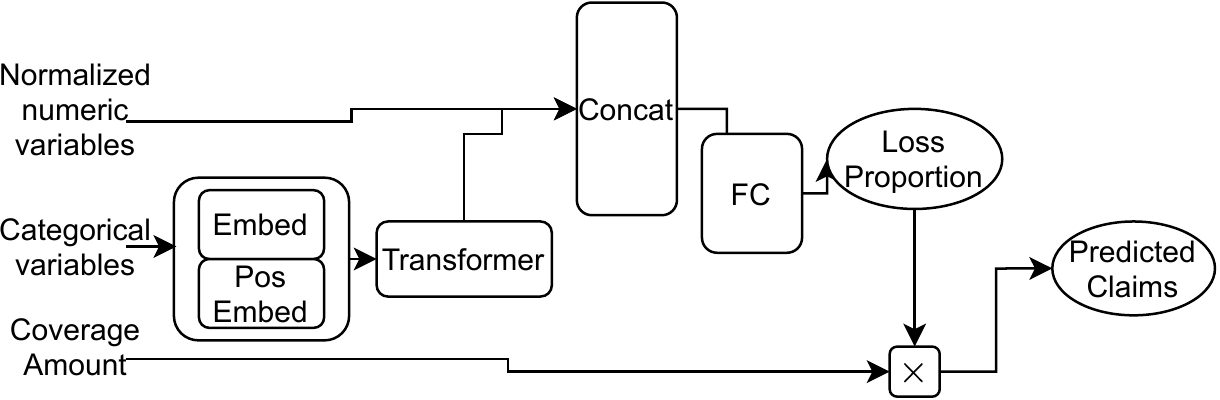} 

}

\caption{Architecture of Model 6.}\label{fig:nn-3}
\end{figure}

Whereas Model 5 makes a simple modification to the structure of Model 4, the TabTransformer model is slightly more complex. With respect to similarities between Model 5 and 6, both use embeddings with a standard 8 dimensions for each categorical covariate. In terms of differences, Model 6 uses a Transformer layer in place of self-attention and a positional encoding of a single dimension, i.e.~\(d_{col} = 1\). The TabTransformer architecture is shown in Figure \ref{fig:nn-3}.

\hypertarget{model-7-tabnet}{%
\subsubsection{Model 7: TabNet}\label{model-7-tabnet}}

The final model we consider in this subsection is the TabNet architecture. In TabNet, similar to decision trees which
it is inspired by, the input features are processed
sequentually in a number of decision steps. In each decision step, a different subset of features are weighted more
than the rest based on masks defined by attention modules applied to information from the previous step. The output
of each step consists of information for the next decision step, and also an output representation that is set aside
then aggregated at the end, along with other decision step outputs, in order to product the final model output.
Due to the complexity, we refer the reader to Arık and Pfister (2019) for the full architectural details.

\hypertarget{training-inference-attn}{%
\subsection{Training and Inference}\label{training-inference-attn}}

We follow a similar training setup for the three attention based models, to the neural networks discussed in the previous section. For Models 5 and 6, we utilize the \texttt{Adam} optimizer as before, however, due to the increased complexity of the models in this section, we use learning rate of 0.001 (compared to 0.01 for the simpler models). Reflecting this reduced learning rate, we train on the analysis set for a maximum of 150 epochs, and select the best model based on the lowest loss achieved on the assessment. Moreover, since these models are significantly more complex than those tested in the previous section, we also add some minimal regularization using dropout (Srivastava et al. 2014) to the attention layers and the last layer of the network, to prevent overfitting. Dropout functions by setting some components of the network randomly to zero during each epoch of training. This forces each component of the network to learn a somewhat independent representation of the data, thus helping to avoid overfitting by preventing components of the network from becoming too specialized. In this section, we drop out components with a probability of 2.5\%. A final detail is that, for both of these models, we use an embedding dimension of size 16 for each of the categorical covariates, since the attention mechanism requires that the embeddings be of a common length. For the TabTransformer model, we use a row embedding of dimension 4.

For Model 7, we use the suggested training setup that utilizes the \texttt{Adam} optimizer with a learning rate of 0.02, and train for 15 epochs (since longer training times did not lead to improved performance).

In Table \ref{tab:cv-results-attn}, we show the cross-validated performance metrics of each of the models, as well as the two best performing models from the previous section. As before, we report both the root mean squared error (RMSE) and the mean absolute error (MAE). The results show that Model 5, which simply adds attention to Model 4, outperforms Model 4 on both the RMSE and MAE metrics. Thus, there is some room to argue that for tabular data modeling that uses embeddings for categorical data, the use of attention within models should be considered. The TabNet architecture performs poorly, perhaps due to insufficient hyperparameter tuning, but indicating that \emph{out of the box}, TabNet does not appear to work well for this modeling task. Finally, the TabTransformer model shows the best performance of all models considered in this research, beating the other models by a significant margin on both the RMSE and MAE metrics.

\begin{longtable}[]{@{}lrr@{}}
\caption{\label{tab:cv-results-attn} Cross validation results for Models 5-7.}\tabularnewline
\toprule
Model & RMSE & MAE\tabularnewline
\midrule
\endfirsthead
\toprule
Model & RMSE & MAE\tabularnewline
\midrule
\endhead
Model 4: MLP with multidimensional embeddings & 60,973 & 37,574\tabularnewline
Model 5: Simple Attention & 60,601 & 36,739\tabularnewline
Model 6: TabNet & 62,938 & 38,386\tabularnewline
Model 7: TabTransformer & 59,900 & 36,343\tabularnewline
Linear regression, predictions capped below at 0.01 & 60,879 & 38,431\tabularnewline
\bottomrule
\end{longtable}

In what follows, we focus on understanding the TabTransformer model in more detail.

\hypertarget{exploring-the-tabtransformer-model}{%
\subsection{Exploring the TabTransformer model}\label{exploring-the-tabtransformer-model}}

Here, we explore what the TabTransformer model has learned from the data via the attention process that forms the core of the model. We focus on the difference between the flood embedding learned by the model, compared with the output of the Transformer layer that processes the embeddings before these enter the model. Huang et al. (2020) call the latter \emph{contextual embeddings}, because the attention mechanism within the TabTransformer model augments the embeddings based on the other covariates that appear for each observation \(i\).

\begin{figure}

{\centering \includegraphics[width=0.8\linewidth,]{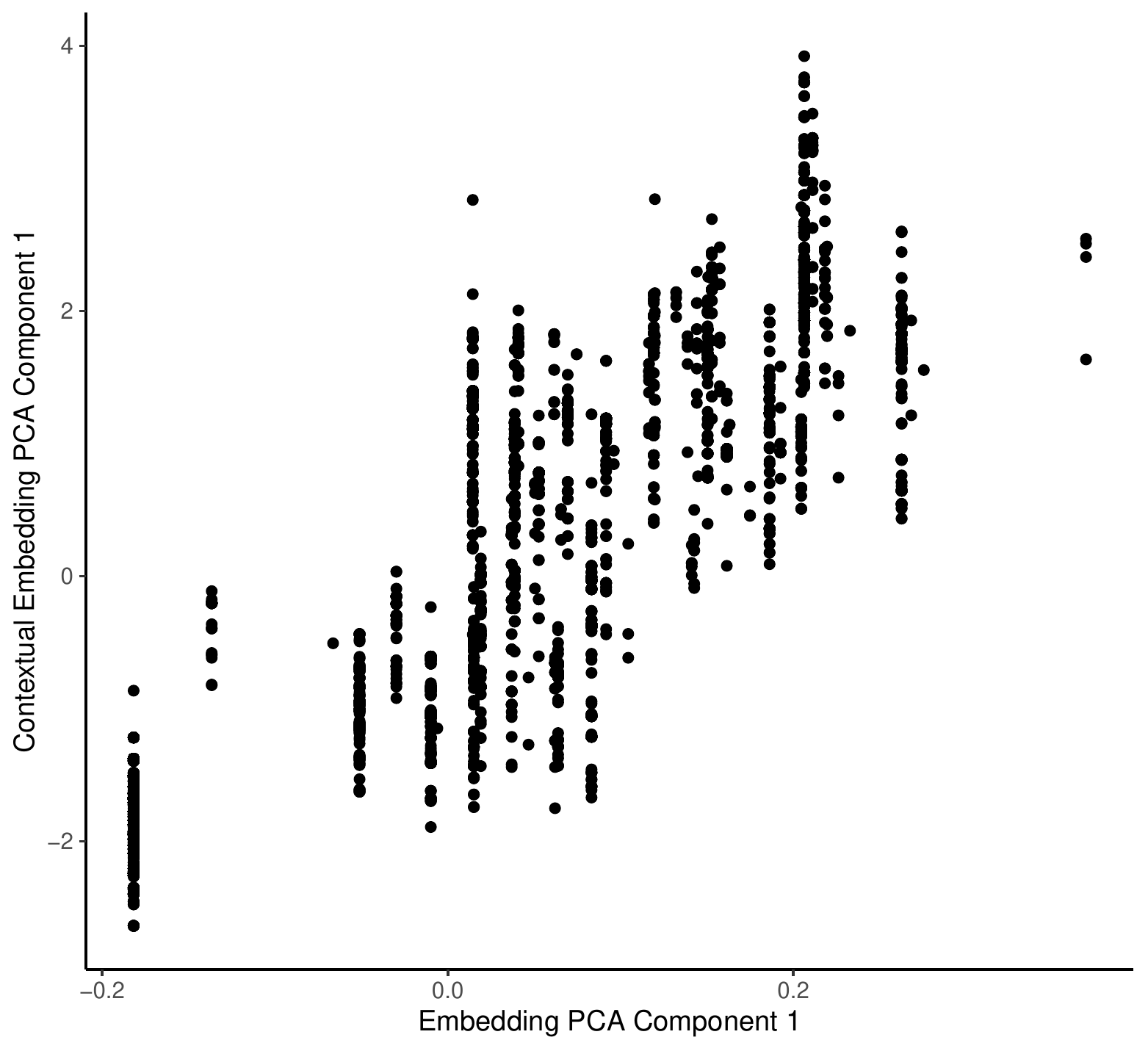} 

}

\caption{Embedding versus Contextual Embedding, Flood Zone, first PCA components}\label{fig:embd}
\end{figure}

To illustrate this, we selected a single TabTransformer fit on one fold of the training data, and selected at random 10 000 individual observations. For each observation, the embedding relating to the flood zone of the observation was extracted from the model, as well as the contextual embedding that was produced for that observation by the TabTransformer model. We found that a PCA and t-SNE analysis of the embeddings produced very similar results to those shown in the previous section, so we do not show these plots here. To investigate the difference between the embeddings and contextual embeddings for the \texttt{flood\_zone} variable, we show the first component from a PCA analysis of each of these in Figure \ref{fig:embd}.

\begin{figure}

{\centering \includegraphics[width=0.8\linewidth,]{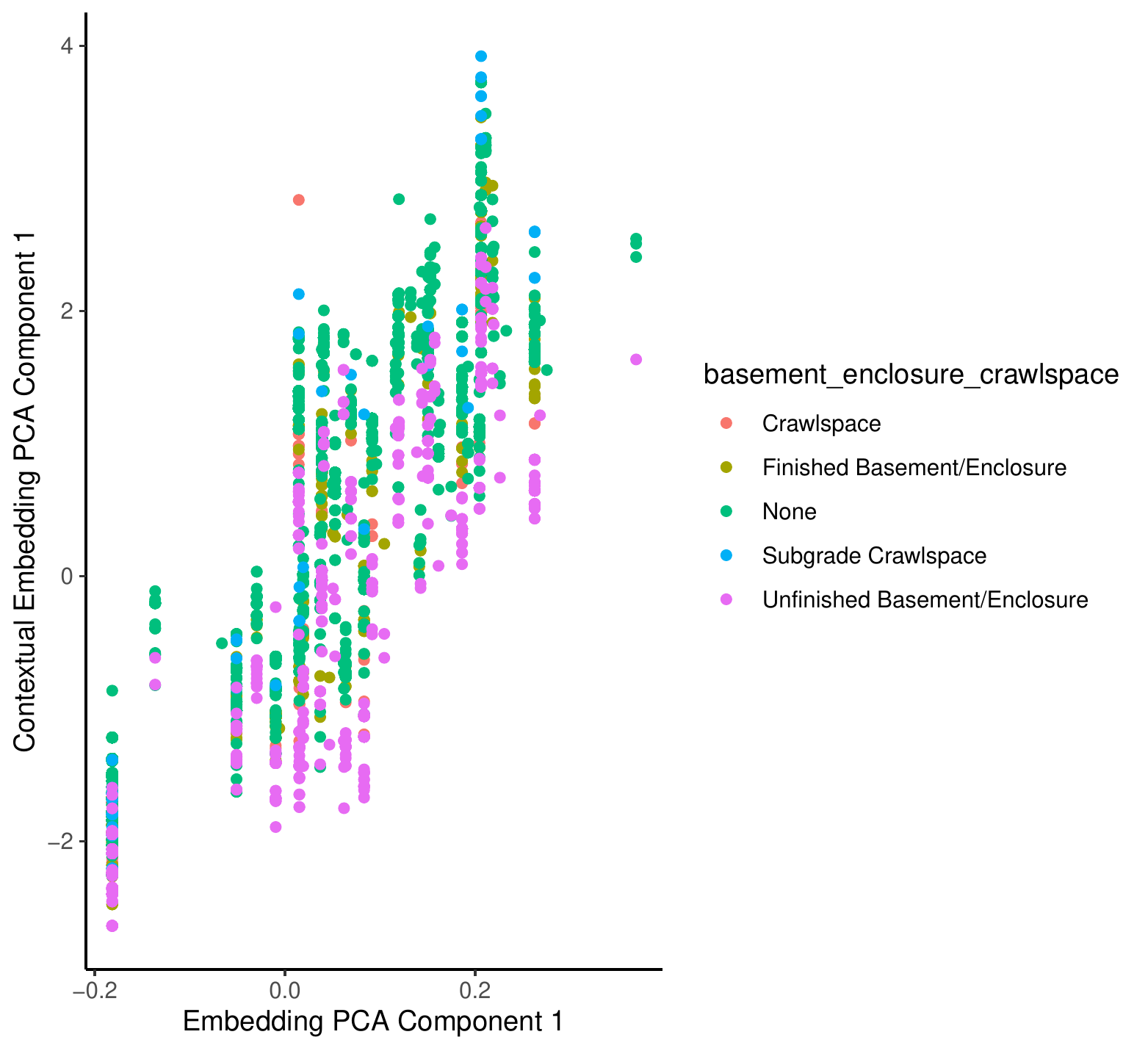} 

}

\caption{Embedding versus Contextual Embedding, Flood Zone, first PCA components, points coloured based on crawl space variable}\label{fig:embdexp}
\end{figure}

The figure shows that, whereas the first principal component of the embeddings occupy discrete points only (shown on the \(x\)-axis), the first principal component of the \textbf{contextual} embeddings is continuous (shown on the \(y\)-axis), meaning that the model has modified the values of the embedding in a smooth manner. Since the inputs to the attention mechanism which is responsible for this are the embeddings for the other categorical variables for each observation, these other variables have influenced the final embedding used within the model for the \texttt{flood\_zone} variable. By considering each of the other categorical variables entering the model, it was found that the basement enclosure crawl space variable explains most of the variation observed within the flood zone embeddings, as shown in Figure \ref{fig:embdexp}. The attention mechanism produces the lowest values for those observations with an unfinished basement, whereas the model produces the highest values for buildings without a crawl space.

To investigate the predictive value added by the contextual embeddings, a simple Gamma GLM was fit to these 10 000 observations, with the goal of predicting the claim amount based only on either the embeddings or the contextual embeddings. As shown in Table \ref{tab:emd-vs-cembd}, the performance of the latter model, as indicated by the RMSE and MAE metrics is better, in other words, the contextual embeddings that incorporate information across all of the categorical variables are better predictors of claim severity compared with the embeddings.

\begin{longtable}[]{@{}lrr@{}}
\caption{\label{tab:emd-vs-cembd} RMSE and MAE for GLMs predicting claim severity, embeddings compared with contextual embeddings.}\tabularnewline
\toprule
Embedding Type & RMSE & MAE\tabularnewline
\midrule
\endfirsthead
\toprule
Embedding Type & RMSE & MAE\tabularnewline
\midrule
\endhead
Embedding & 62596.84 & 40323.82\tabularnewline
Contextual embedding & 62027.85 & 39450.45\tabularnewline
\bottomrule
\end{longtable}

\newpage

\hypertarget{Conclusions}{%
\section{Conclusions}\label{Conclusions}}

This study has focused on how embeddings can be applied within the context of actuarial modeling, by illustrating a range of approaches: first, relatively simple neural networks with embeddings were defined, the embeddings from these models were reused within traditional GLM models and, finally, more advanced attention based models were studied. These models were demonstrated within the context of a claims severity modeling problem based on data from the NFIP. Whereas we have only considered severity in this study, it is likely that our findings will replicate for the modeling of claim frequency, which is usually an easier problem. Our key findings are that processing categorical variables with multi-dimensional embeddings leads to enhanced performance within simple neural networks, and that, by including these within traditional GLM models, the predictive performance of these models can be enhanced, compared to the usual encoding scheme for categorical variables. Some of the attention based models considered within this study performed well, in particular, the TabTransformer model of Huang et al. (2020) appears quite promising for use in actuarial modeling. The study has also illustrated how embeddings can be visualized using the PCA and t-SNE techniques. We found that, in the case of the NFIP data, it is possible to explain which variables modify the embeddings based on the context of each observation.

Several avenues for future research can be considered. One surprising finding of our work is that the TabNet model performed quite poorly, leading us to conclude that this model appears to be less suitable for actuarial modeling. Understanding these findings in more detail would be a valuable contribution to the literature. Another helpful contribution would be to investigate the optimal hyperparameters for the TabTransformer model in the context of actuarial modeling. Finally, extending our findings to a claims frequency example could be considered.

\clearpage

\appendix

\hypertarget{appendix}{%
\section*{Appendix}\label{appendix}}
\addcontentsline{toc}{section}{Appendix}

\hypertarget{EDA}{%
\section{Exploratory Data Analysis}\label{EDA}}

\begin{figure}

{\centering \includegraphics[width=0.6\linewidth,]{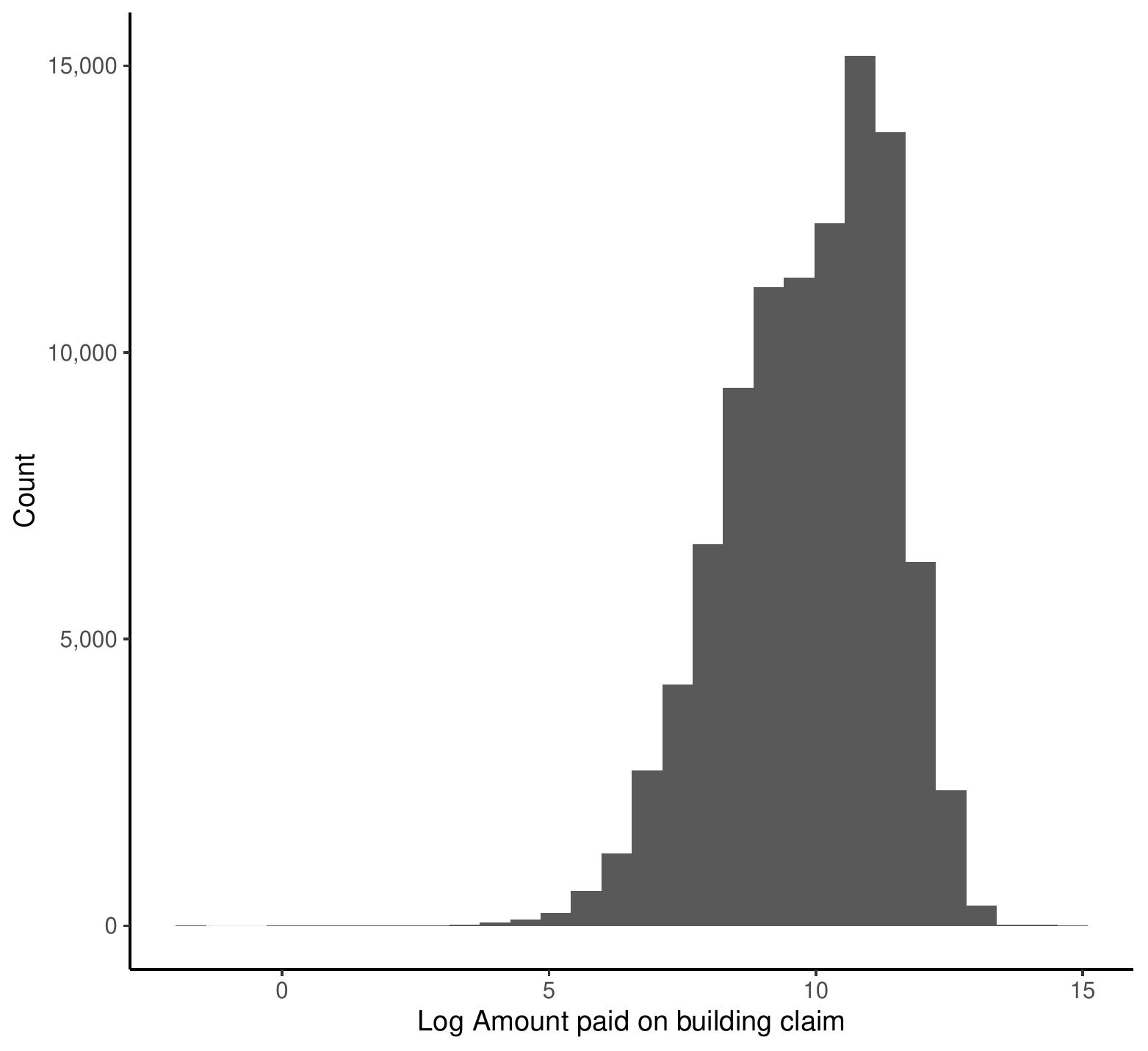} 

}

\caption{Histogram of the Log Claim Amount}\label{fig:sev}
\end{figure}

\begin{figure}

{\centering \includegraphics[width=0.6\linewidth,]{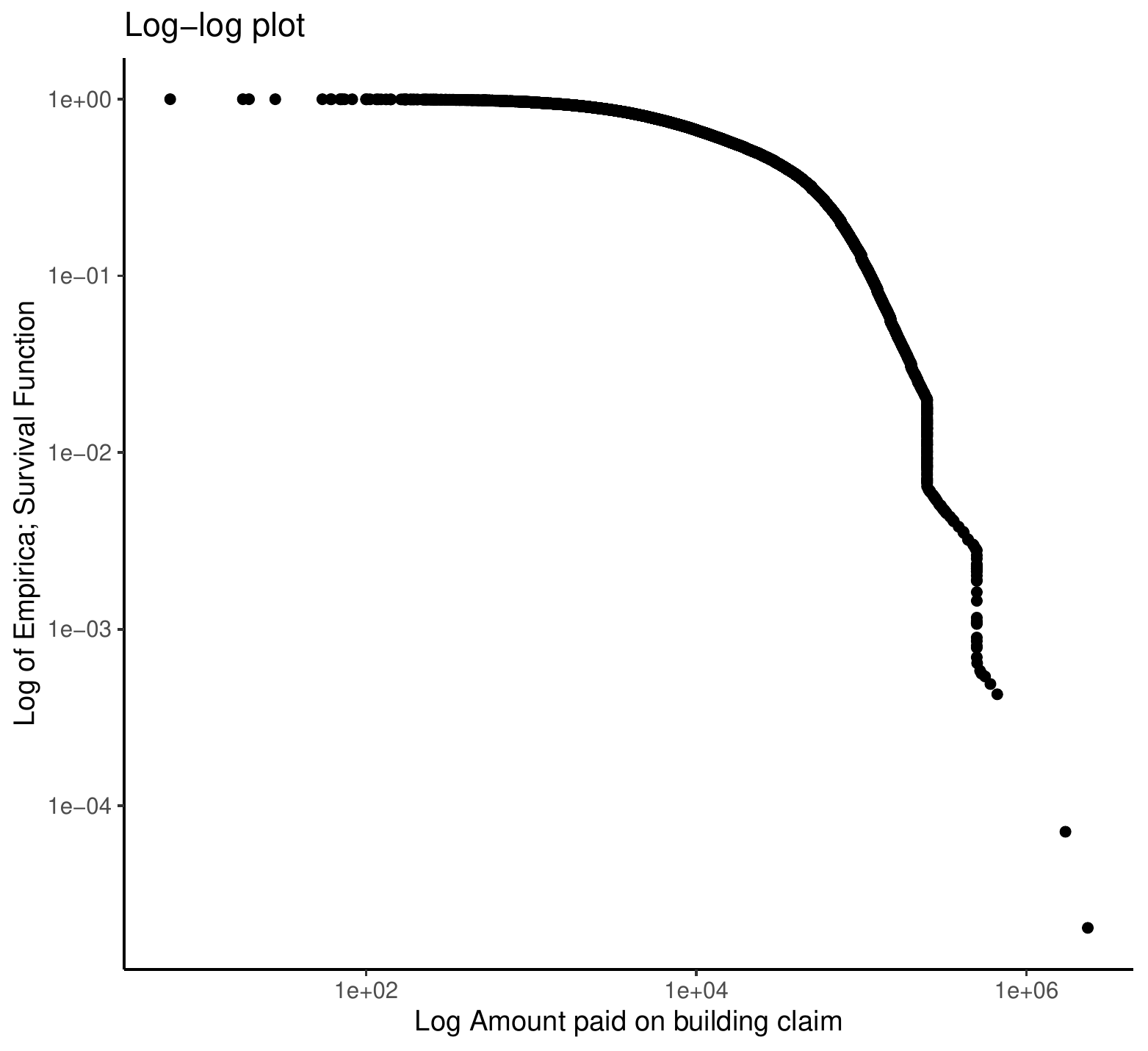} 

}

\caption{Log-log plot of Claim amount}\label{fig:loglog}
\end{figure}

\begin{figure}

{\centering \includegraphics[width=0.6\linewidth,]{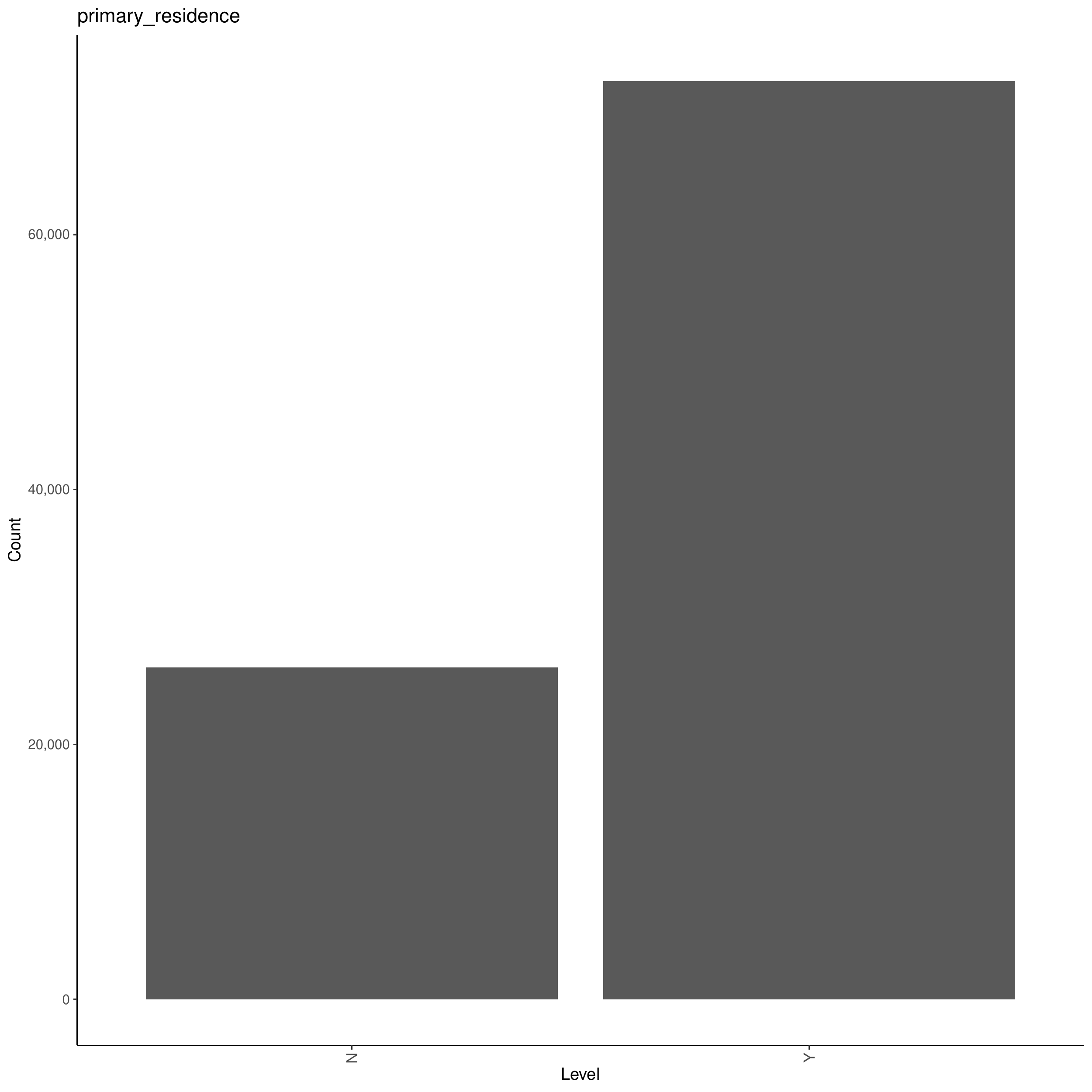} 

}

\caption{Distribution of observations across the Primary Residence covariate}\label{fig:catprimres}
\end{figure}

\begin{figure}

{\centering \includegraphics[width=0.6\linewidth,]{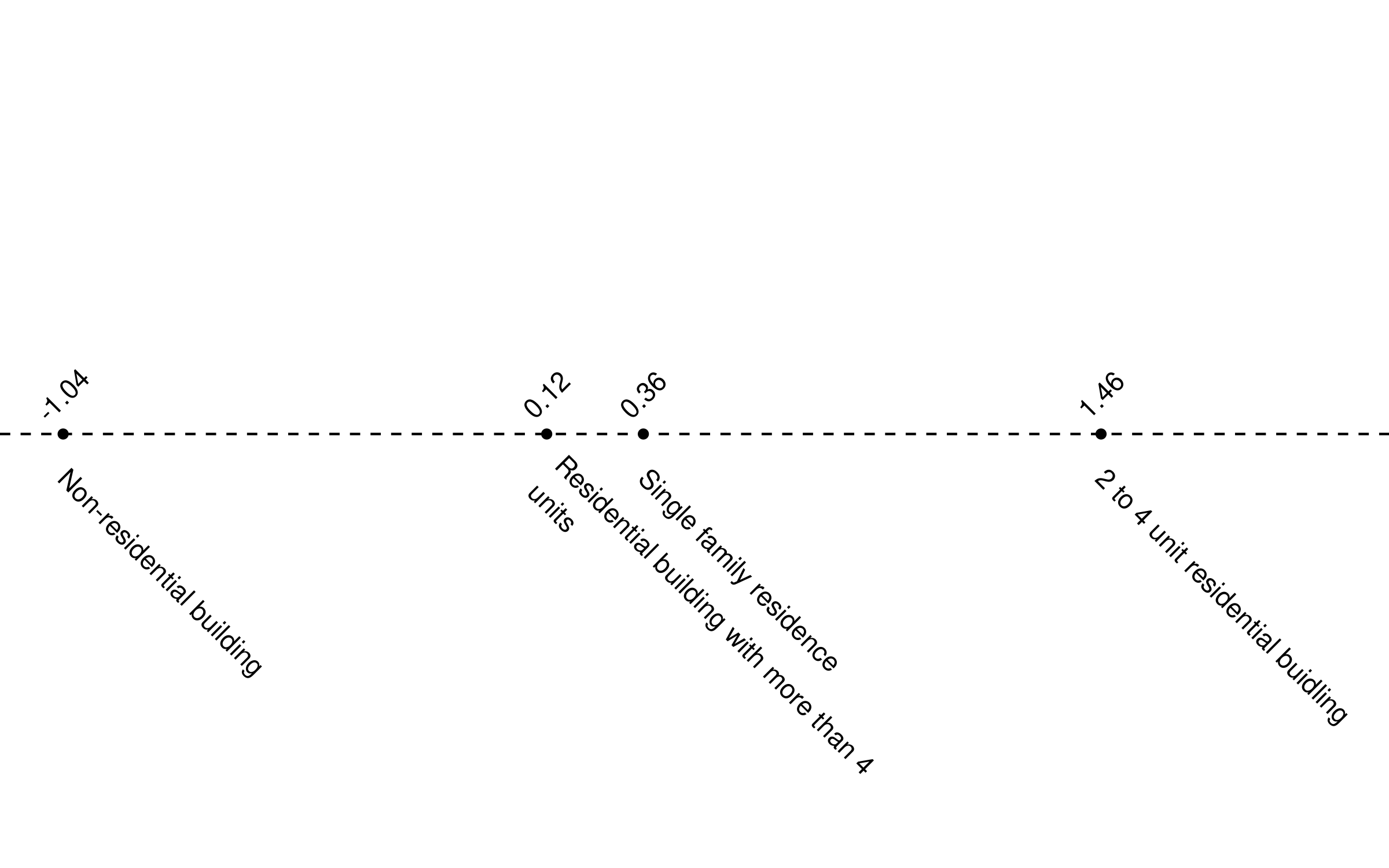} 

}

\caption{Distribution of observations across the Occupancy type covariate}\label{fig:catocc}
\end{figure}

\begin{figure}

{\centering \includegraphics[width=0.6\linewidth,]{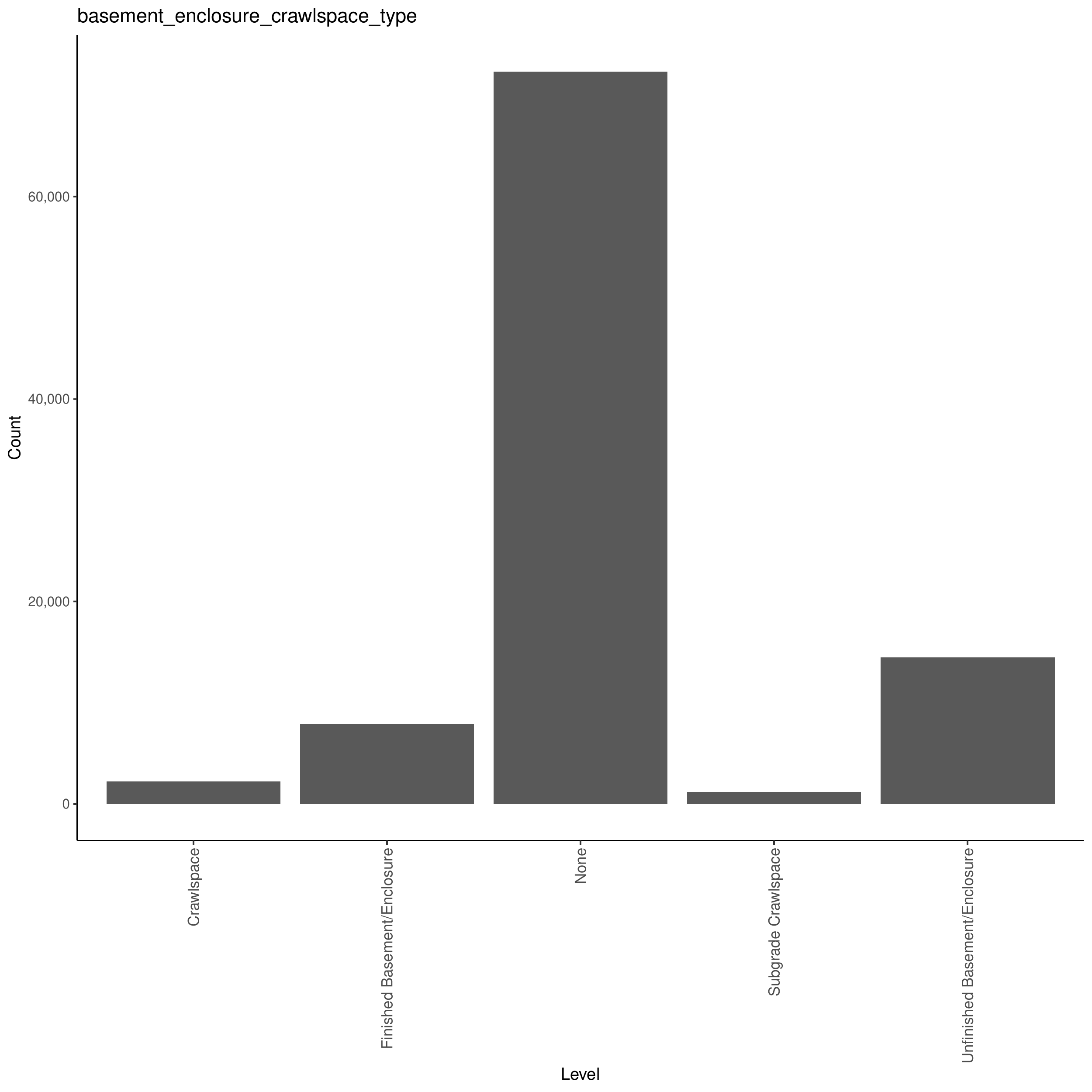} 

}

\caption{Distribution of observations across the Basement Enclosure and Crawlspace covariate}\label{fig:catbasement}
\end{figure}

\begin{figure}

{\centering \includegraphics[width=0.6\linewidth,]{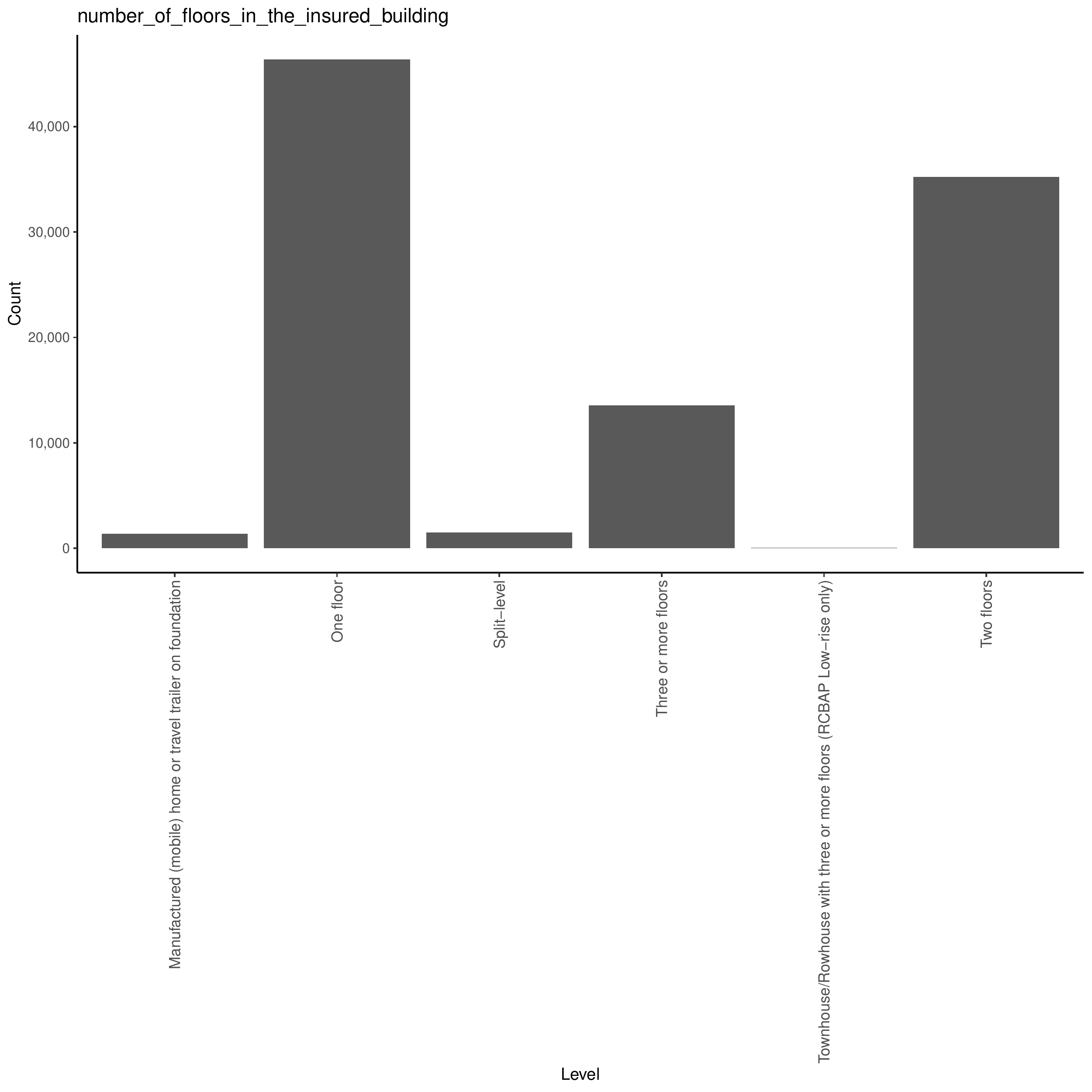} 

}

\caption{Distribution of observations across the Number of Floors covariate}\label{fig:catfloors}
\end{figure}

\begin{figure}

{\centering \includegraphics[width=0.6\linewidth,]{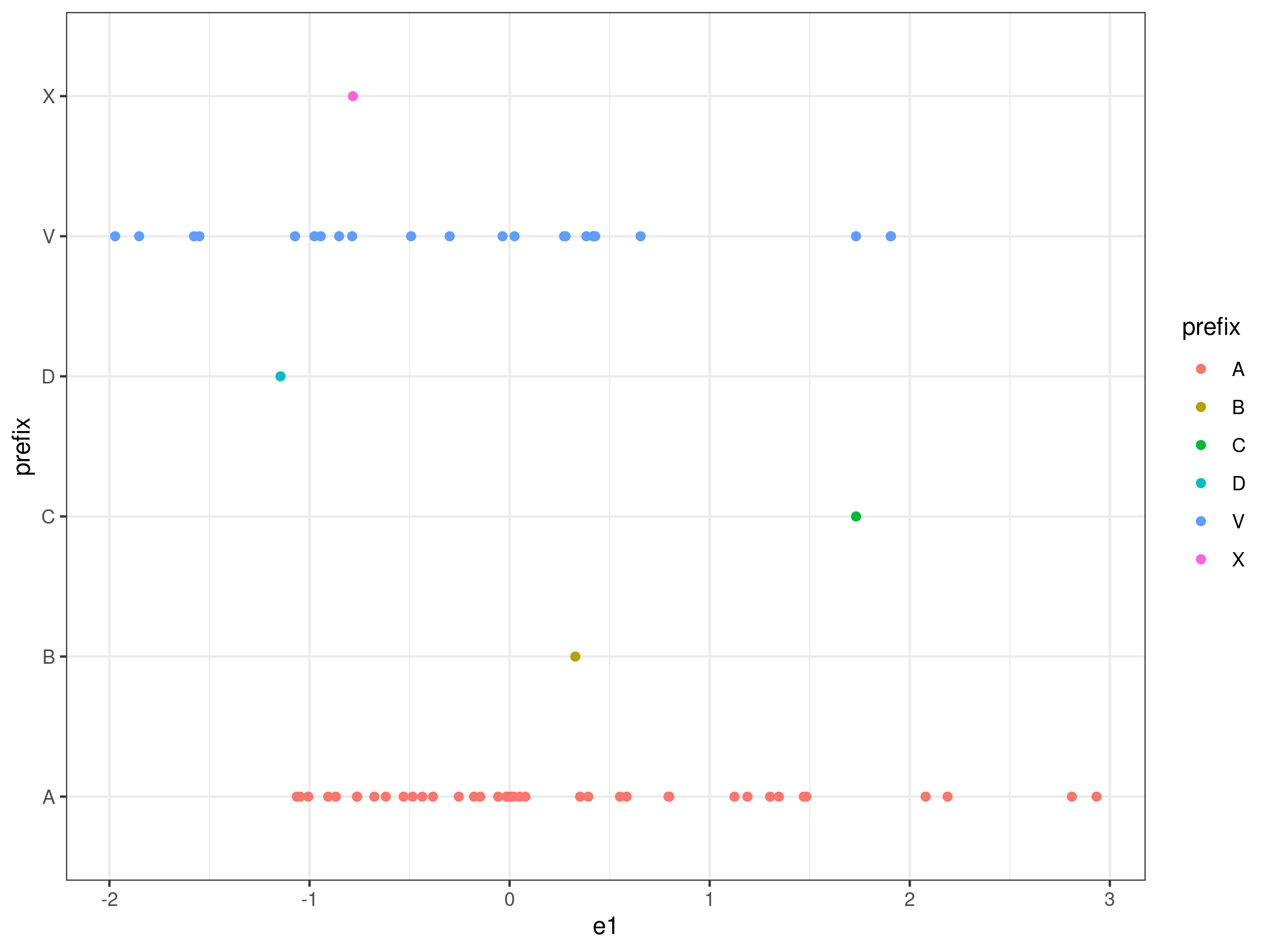} 

}

\caption{Distribution of observations across the Flood Zone covariate}\label{fig:catflood}
\end{figure}

\begin{figure}

{\centering \includegraphics[width=0.6\linewidth,]{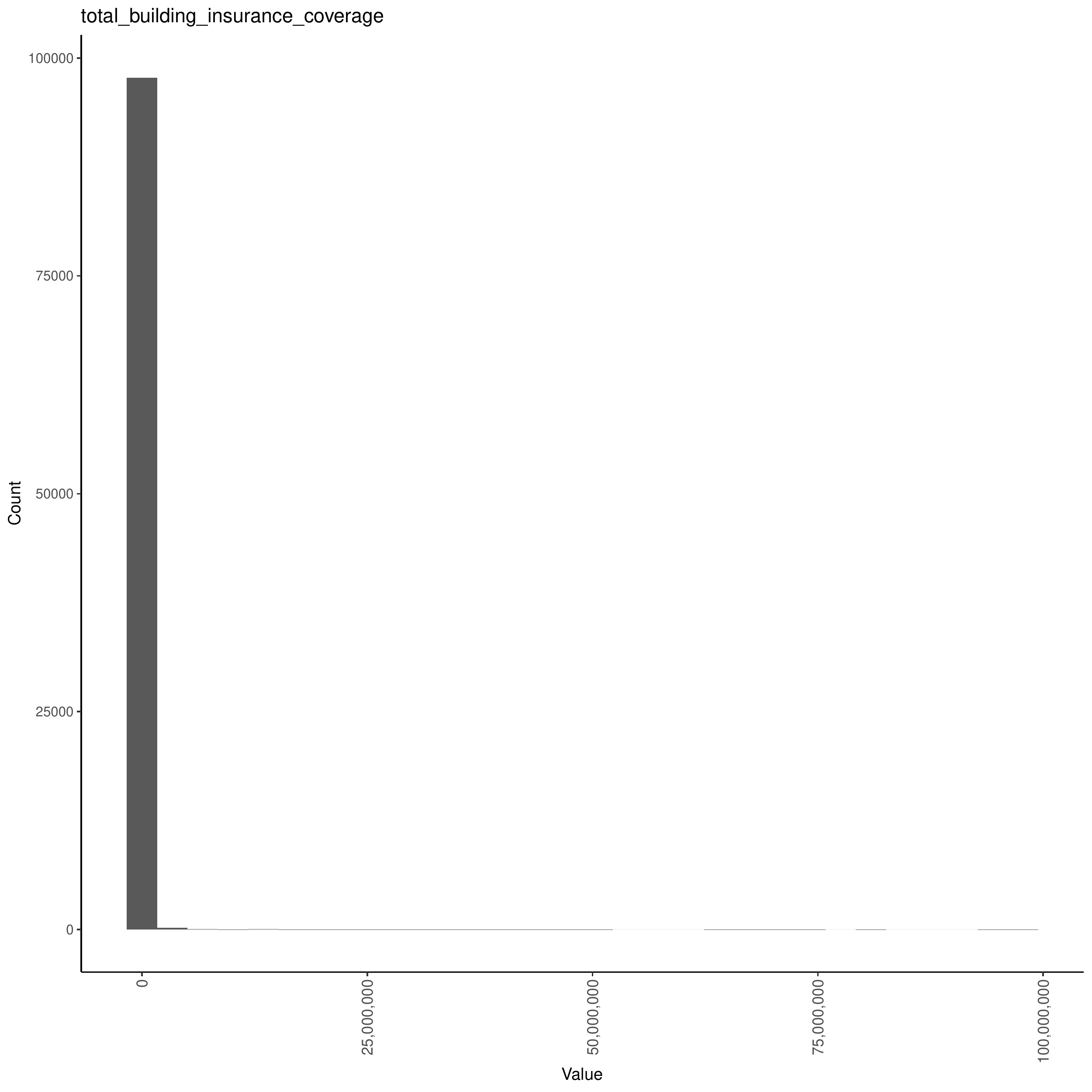} 

}

\caption{Histogram of the Building Coverage covariate}\label{fig:numcov}
\end{figure}

\begin{figure}

{\centering \includegraphics[width=0.6\linewidth,]{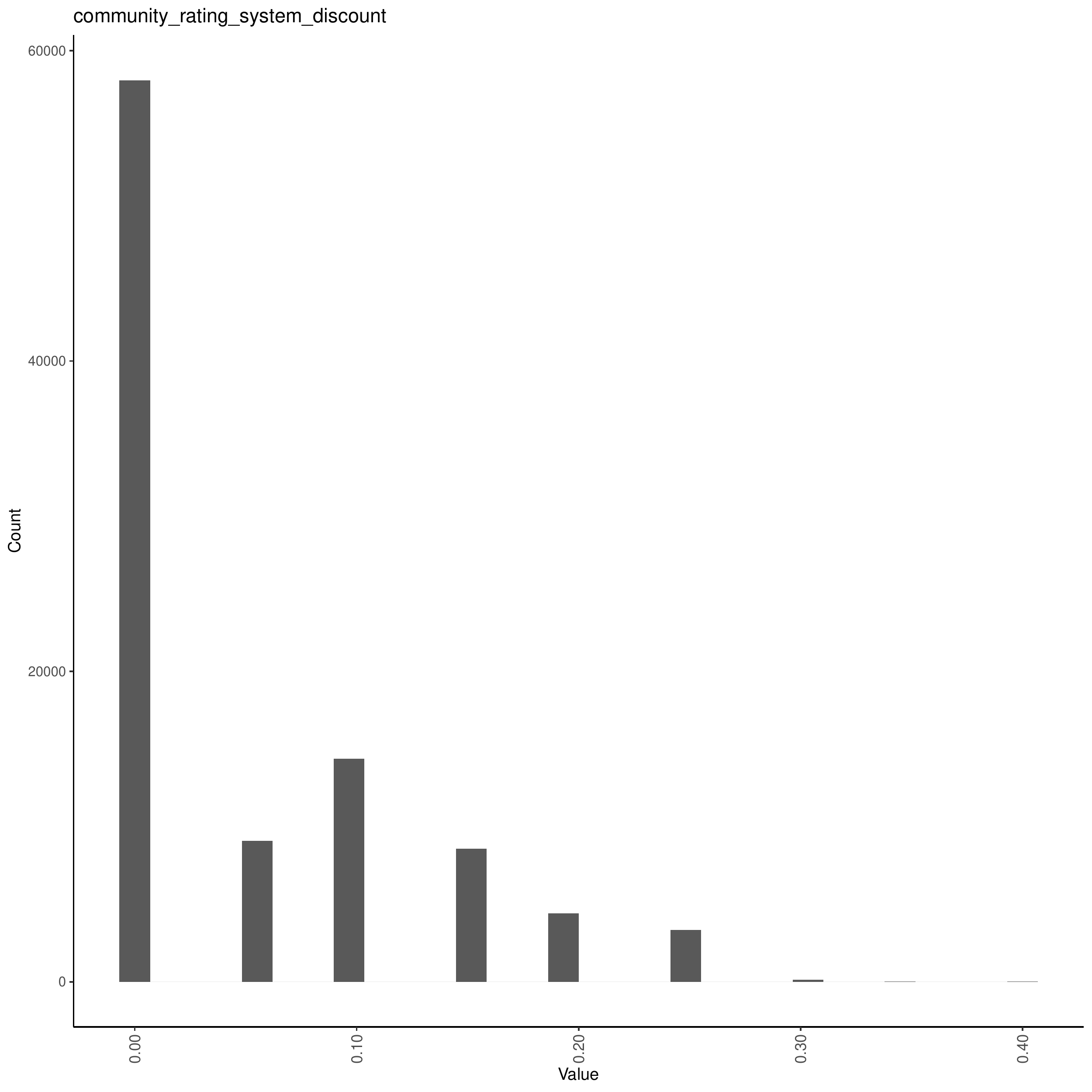} 

}

\caption{Distribution of the Comunity Rating System Discount covariate}\label{fig:numdisc}
\end{figure}

In this appendix we present an exploratory data analysis of the NFIP claims dataset. Figure \ref{fig:sev} shows the distribution of the claim size that we model in this study. It can be seen that claim size varies over several orders of magnitude. Figure \ref{fig:loglog} shows the claim size seemingly follows a relatively heavy-tailed distribution. Figures \ref{fig:catprimres} - \ref{fig:catflood} illustrate the distribution of the categorical covariates within the dataset. Similarly, Figures \ref{fig:numcov} - \ref{fig:numdisc} illustrate the distribution of the continuous covariates within the dataset.

\clearpage

\hypertarget{references}{%
\section*{References}\label{references}}
\addcontentsline{toc}{section}{References}

\hypertarget{refs}{}
\begin{CSLReferences}{1}{0}
\leavevmode\hypertarget{ref-Ark2019}{}%
Arık, Sercan, and Tomas Pfister. 2019. {``{TabNet: Attentive Interpretable Ta bular Learning}.''} \emph{arXiv}, August. \href{https://www.aaai.org}{www.aaai.org}.

\leavevmode\hypertarget{ref-Bahdanau2015a}{}%
Bahdanau, Dzmitry, Kyung Hyun Cho, and Yoshua Bengio. 2015. {``{Neural machine translation by jointly learning to align and translate}.''} \emph{3rd International Conference on Learning Representations, ICLR 2015 - Conference Track Proceedings}, 1--15. \url{http://arxiv.org/abs/1409.0473}.

\leavevmode\hypertarget{ref-Buhlmann2006}{}%
Bühlmann, Hans, and Alois Gisler. 2005. \emph{{A Course in Credibility Theory and its Applications}}. Springer Science {\&} Business Media. \url{https://doi.org/10.1080/03461230600889660}.

\leavevmode\hypertarget{ref-Delong2020a}{}%
Delong, Lukasz, Mathias Lindholm, and Mario V. Wuthrich. 2020. {``{Collective Reserving using Individual Claims Data}.''} \emph{SSRN Electronic Journal}, May. \url{https://doi.org/10.2139/ssrn.3582398}.

\leavevmode\hypertarget{ref-FederalEmergencyManagementAgency2019}{}%
Federal Emergency Management Agency. 2019. {``{FIMA NFIP Redacted Claims Data Set}.''} \url{https://www.fema.gov/media-library/assets/documents/180376\%20https://www.fema.gov/media-library/assets/documents/180374}.

\leavevmode\hypertarget{ref-Gabrielli2019c}{}%
Gabrielli, Andrea. 2019. {``{A Neural Network Boosted Double Over-Dispersed Poisson Claims Reserving Model}.''} \emph{SSRN Electronic Journal}, April. \url{https://doi.org/10.2139/ssrn.3365517}.

\leavevmode\hypertarget{ref-Gabrielli2019}{}%
Gabrielli, Andrea, Ronald Richman, and Mario V. Wüthrich. 2019. {``{Neural network embedding of the over-dispersed Poisson reserving model}.''} \emph{Scandinavian Actuarial Journal}. \url{https://doi.org/10.1080/03461238.2019.1633394}.

\leavevmode\hypertarget{ref-Goldburd}{}%
Goldburd, Mark, Anand Khare, Dan Tevet, and Dmitriy Guller. 2020. \emph{{Generalized Linear Models for Insurance Rating}}. Second Edi. Casualty Actuarial Society. \href{https://www.casact.org}{www.casact.org}.

\leavevmode\hypertarget{ref-Guo2016}{}%
Guo, Cheng, and Felix Berkhahn. 2016. {``{Entity Embeddings of Categorical Variables}.''} \emph{arXiv} arXiv:1604. \url{http://arxiv.org/abs/1604.06737}.

\leavevmode\hypertarget{ref-henckaerts2020boosting}{}%
Henckaerts, Roel, Marie-Pier Côté, Katrien Antonio, and Roel Verbelen. 2020. {``Boosting Insights in Insurance Tariff Plans with Tree-Based Machine Learning Methods.''} \emph{North American Actuarial Journal}, 1--31.

\leavevmode\hypertarget{ref-Huang2020}{}%
Huang, Xin, Ashish Khetan, Milan Cvitkovic, and Zohar Karnin. 2020. {``{TabTransformer: Tabular data modeling using contextual embeddings}.''} \emph{arXiv}, December. \url{http://arxiv.org/abs/2012.06678}.

\leavevmode\hypertarget{ref-Kingma2014}{}%
Kingma, Diederik P., and Jimmy Ba. 2014. {``{Adam: A Method for Stochastic Optimization}.''} \emph{arXiv Preprint arXiv:1412.6980}. \url{http://arxiv.org/abs/1412.6980}.

\leavevmode\hypertarget{ref-Klinker2010}{}%
Klinker, Fred. 2010. {``{Generalized Linear Mixed Models for Ratemaking: A Means of Introducing Credibility into a Generalized Linear Model Setting}.''} \emph{Casualty Actuarial Society E-Forum, Winter 2011 Volume 2} 2 (1): 1--25. \url{http://scholar.google.com/scholar?hl=en\&btnG=Search\&q=intitle:Generalized+Linear+Mixed+Models+for+Ratemaking+:+A+Means+of+Introducing+Credibility+into+a+Generalized+Linear+Model+Setting\#0}.

\leavevmode\hypertarget{ref-Kuo2019}{}%
Kuo, Kevin. 2019. {``{Deeptriangle: A deep learning approach to loss reserving}.''} \emph{Risks} 7 (3). \url{https://doi.org/10.3390/risks7030097}.

\leavevmode\hypertarget{ref-Kuo2020}{}%
---------. 2020. {``{Individual Claims Forecasting with Bayesian Mixture Density Networks},''} March. \url{http://arxiv.org/abs/2003.02453}.

\leavevmode\hypertarget{ref-kuo2020explainability}{}%
Kuo, Kevin, and Daniel Lupton. 2020. {``Towards Explainability of Machine Learning Models in Insurance Pricing.''} \url{http://arxiv.org/abs/2003.10674}.

\leavevmode\hypertarget{ref-Maaten2008a}{}%
Maaten, L, and G Hinton. 2008. {``{Visualizing data using t-SNE}.''} \emph{Journal of Machine Learning Research} 9 (Nov): 2579--2605.

\leavevmode\hypertarget{ref-Richman2020}{}%
Perla, Francesca, Ronald Richman, Salvatore Scognamiglio, and Mario V. Wüthrich. 2020. {``{Time-Series Forecasting of Mortality Rates using Deep Learning}.''} \emph{SSRN Electronic Journal}.

\leavevmode\hypertarget{ref-Richman2018}{}%
Richman, R. 2018. {``{AI in Actuarial Science}.''} \emph{SSRN Electronic Journal}, October. \url{https://doi.org/10.2139/ssrn.3218082}.

\leavevmode\hypertarget{ref-Richman2019d}{}%
Richman, R, and Mario V. Wüthrich. 2019. {``{A neural network extension of the Lee-Carter model to multiple populations}.''} \emph{Annals of Actuarial Science}. \url{https://doi.org/10.1017/S1748499519000071}.

\leavevmode\hypertarget{ref-schelldorfer2019nesting}{}%
Schelldorfer, Jürg, and Mario V. Wüthrich. 2019. {``{Nesting Classical Actuarial Models into Neural Networks}.''} \emph{SSRN Electronic Journal}. \url{https://doi.org/10.2139/ssrn.3320525}.

\leavevmode\hypertarget{ref-shlens2014tutorial}{}%
Shlens, Jonathon. 2014. {``A Tutorial on Principal Component Analysis.''} \emph{arXiv Preprint arXiv:1404.1100}.

\leavevmode\hypertarget{ref-srivastava2014dropout}{}%
Srivastava, Nitish, Geoffrey Hinton, Alex Krizhevsky, Ilya Sutskever, and Ruslan Salakhutdinov. 2014. {``{Dropout: A simple way to prevent neural networks from overfitting}.''} \emph{Journal of Machine Learning Research} 15 (1): 1929--58.

\leavevmode\hypertarget{ref-Vaswani2017a}{}%
Vaswani, Ashish, Noam Shazeer, Niki Parmar, Jakob Uszkoreit, Llion Jones, Aidan N Gomez, Łukasz Kaiser, and Illia Polosukhin. 2017b. {``{Attention is all you need}.''} In \emph{Advances in Neural Information Processing Systems}, 2017-Decem:5999--6009. \url{http://arxiv.org/abs/1706.03762v5}.

\leavevmode\hypertarget{ref-Vaswani2017}{}%
---------. 2017a. {``{Attention is all you need}.''} In \emph{Advances in Neural Information Processing Systems}, 2017-Decem:5999--6009. \url{http://arxiv.org/abs/1706.03762v5}.

\leavevmode\hypertarget{ref-wattenberg2016how}{}%
Wattenberg, Martin, Fernanda Viégas, and Ian Johnson. 2016. {``How to Use t-SNE Effectively.''} \emph{Distill}. \url{https://doi.org/10.23915/distill.00002}.

\leavevmode\hypertarget{ref-wuthrich2019yes}{}%
Wüthrich, Mario V, and Michael Merz. 2019. {``{Yes, we CANN!}''} \emph{ASTIN Bulletin: The Journal of the IAA} 49 (1): 1--3.

\end{CSLReferences}

\bibliographystyle{unsrt}
\bibliography{references.bib}

\end{document}